\documentclass[11pt]{article}
\usepackage{subfigure}
\usepackage{graphics}
\usepackage{amsmath}
\usepackage{amssymb}
\usepackage{epsf}
\usepackage{epsfig}
\usepackage{rotating}
\usepackage{pstricks}
\usepackage{color}
\usepackage{url}

\def\ds#1{#1\kern-1ex\hbox{/}}
\def\dsh{h\kern-1.2ex /}

\newcommand{\bea}{\begin{eqnarray}}
\newcommand{\eea}{\end{eqnarray}}

\def\beq{\begin{equation}}
\def\eeq{\end{equation}}

\def\beqn{\begin{eqnarray}}
\def\eeqn{\end{eqnarray}}
\def\ba{\begin{eqnarray}}
\def\ea{\end{eqnarray}}

\newcommand{\be}{\begin{equation}}
\newcommand{\beqa}{\begin{eqnarray}}
\newcommand{\eeqa}{\end{eqnarray}}

\newcommand{\ee}{\end{equation}}

\begin{document}
\begin{center}
\vspace{5.cm}

{\bf \large Nonperturbative contributions to a resummed leptonic angular
distribution in inclusive neutral vector boson production}
\vspace{0.5cm}

\vspace{1cm}
{ \bf Marco Guzzi$^{a}$, Pavel M. Nadolsky$^{b}$ and Bowen Wang$^{b}$\\}


{\it {$^a$}Deutsches Elektronen-Synchrotron DESY,\\
Notkestrasse 85, 22607 Hamburg, Germany\\
{$^b$}Department of Physics, Southern Methodist University, \\Dallas, TX 75275, USA}
\vspace{.5cm}\footnote{marco.guzzi@desy.de, nadolsky@physics.smu.edu, bwang@physics.smu.edu}

\begin{abstract}

We present an analysis of nonperturbative contributions to 
the transverse momentum distribution of $Z/\gamma^*$ bosons 
produced at hadron colliders. The new data on the angular 
distribution $\phi^*_\eta$ of Drell-Yan pairs 
measured at the Tevatron is shown to be in excellent agreement 
with a perturbative QCD prediction based 
on the Collins-Soper-Sterman (CSS) resummation formalism
at NNLL accuracy. Using these data, we  determine the nonperturbative
component of the CSS resummed cross section and estimate its dependence
on arbitrary resummation scales and other factors.
With the scale dependence included at the NNLL level, 
a significant nonperturbative component is needed to describe the angular data.
\end{abstract}
\end{center}
\date{\today}
\newpage

\section{Introduction}
QCD factorization methods utilizing transverse-momentum-dependent (TMD)
parton distributions and fragmentation functions provide a powerful
framework for describing multi-scale observables in high-energy
hadron interactions. Production of Drell-Yan lepton-antilepton 
pairs in $Z/\gamma^*$ boson production in hadron-hadron collisions 
is one basic process in which TMD factorization is applied 
to predict the boson's transverse momentum ($Q_T$) distribution 
and related angular distributions. 
Collinear QCD factorization is applicable for describing
lepton pairs with $Q_T$ of order of the
invariant mass $Q$ of the pair. The respective large-$Q_T$ cross sections have
been computed up to two loops in the QCD coupling strength
$\alpha_s$~\cite{Arnold:1988dp,Anastasiou:2003ds,Catani:2009sm,Bozzi:2010xn}
and are in reasonable agreement with the data.  

But, at small $Q_T$, all-order resummation of large logarithms $\ln(Q_T/Q)$
needs to be performed~\cite{Dokshitzer:1978yd,Parisi:1979se,Curci:1979bg} to obtain sensible
cross sections. TMD factorization provides a systematic framework for 
$Q_T$ resummation to all orders in $\alpha_s$, 
as has been shown in classical papers by Collins, Soper,
and Sterman (CSS)~\cite{Collins:1981uk,Collins:1981va,Collins:1984kg,Collins:2004nx,CollinsBook2}.
The resummed cross sections have been computed at various QCD orders in
the CSS formalism and kindred approaches ~\cite{Balazs:1995nz,Ellis:1997ii,deFlorian:2000pr,Catani:2000vq,deFlorian:2001zd,Catani:2007vq,Bozzi:2008bb,Catani:2009sm,Bozzi:2010xn,Banfi:2012du,Banfi:2011dx,Banfi:2009dy}. 
In addition to perturbative radiative contributions, the resummed
cross sections include a nonperturbative component associated with QCD
dynamics at momentum scales below 1 GeV. Understanding 
of the nonperturbative terms 
is important for tests of TMD factorization and precision studies 
of electroweak boson production, including the measurement 
of $W$ boson mass~\cite{Nadolsky:2004vt}. 

Instead of measuring $Q_T$ distributions directly, one can measure the 
distribution in the angle $\phi^{*}_{\eta}$~\cite{Banfi:2010cf} that is closely related to $Q_T/Q$.
The $\phi^*_\eta$ distributions have been recently measured both at the Tevatron~\cite{Abazov:2010mk} 
and Large Hadron Collider~\cite{Aad:2011gj,Aad:2012wfa}. 
Small experimental errors of the $\phi^*_\eta$ measurements (as low as 0.5\%) 
allow one to test the $Q_T$ resummation formalism at an unprecedented level.
On the theory side, the small-$Q_T$ resummed form factor  
for $Z$ boson production has been computed to NNLL/NNLO~\cite{Catani:2012qa}.\footnote{Throughout the paper, 
``NNLO'' will consistently refer to the cross sections of order $\alpha_s^2$, in accordance with the 
observation that the lowest-order non-zero contribution to the resummed $Q_T$ distribution arises 
from the subprocess $q\bar q \rightarrow V$ of order $\alpha_s^0$. This is to be distinguished from an alternative convention that may be applied 
at large $Q_T$~\cite{Catani:2012qa,Banfi:2012du}, according to which the $\alpha_s^2$ contributions are of the next-to-leading order (NLO).}  
We would like to confront precise theoretical predictions implemented
in programs \textsc{Legacy} and \textsc{ResBos}~\cite{Ladinsky:1993zn,Balazs:1997xd, Landry:2002ix} by the
new experimental data to obtain quantitative constraints on the
nonperturbative contributions. 

Such analysis is technically challenging and requires to examine
several effects that were negligible in the previous studies 
of the resummed nonperturbative terms~\cite{Ladinsky:1993zn,Landry:2002ix,Konychev:2005iy}. The framework
for the fitting of Drell-Yan processes 
in the CSS formalism must be extended to the
$\phi^*_\eta$, rather than $Q_T$, distributions. 
Nonperturbative effects must be distinguished 
from comparable modifications by NNLO QCD 
corrections, NLO electroweak (EW) corrections, 
and the associated perturbative uncertainties.

To carry out this study, we modified the $Q_T$
resummation calculation employed in our previous studies to evaluate NNLO
QCD ($\alpha_s^2$) and NLO EW ($\alpha_{EW}$) perturbative contributions
and consider the residual QCD scale dependence associated with
higher-order terms. This implementation was utilized to determine the
nonperturbative factor from the D\O\ Run-2 data on the
$\phi^*_\eta$ distributions. 

Our findings shed light on several questions raised in recent
studies of TMD factorization~\cite{Collins:1999dz,Collins:2000gd,Henneman:2001ev,Belitsky:2002sm,Boer:2003cm,Collins:2003fm,Collins:2007ph,Idilbi:2005er,Cherednikov:2007tw,Cherednikov:2009wk,GarciaEchevarria:2011rb,Echevarria:2012pw,Aybat:2011zv}
and soft-collinear-effective (SCET) theory~\cite{Becher:2011xn,Mantry:2010mk,Mantry:2010bi}. 
We examine if the $\phi^*_\eta$ data corroborate the universal behavior of the
resummed nonperturbative terms that is expected from the TMD
factorization theorem~\cite{Collins:2004nx} and was observed in the global
analyses of Drell-Yan $Q_T$ distributions at fixed-target and collider energies~\cite{Landry:2002ix,Konychev:2005iy}. 
We also investigate the rapidity dependence of the nonperturbative
terms, which may be indicative of new types 
of higher-order contributions~\cite{Berge:2004nt}.
It has been argued~\cite{Bozzi:2010xn,Banfi:2011dm,Banfi:2012du,Banfi:2011dx,Banfi:2009dy} that
the evidence for nonperturbative smearing is inconclusive 
because of a large QCD scale dependence. Since the magnitude of the
scale dependence reduces with the order of the calculation,  
we include the dependence on the soft scales in the resummed cross section up
to ${\cal O}(\alpha_s^2)$, {\it i.e.}~NNLL/NNLO. In this case, the  
radiative contributions are estimated to the same order as in~\cite{Catani:2012qa}, either exactly or approximately, and we also
include contributions responsible for the dependence on the
resummation scales to one higher order ($\alpha_s^2$) than in~\cite{Bozzi:2010xn,Banfi:2011dm,Banfi:2012du,Banfi:2011dx,Banfi:2009dy}.  

Based on our numerical implementation, we demonstrate 
that the impact of the power-suppressed contributions
is generically distinct from the scale dependence:
the nonperturbative effects can be distinguished 
from the NNLO scale uncertainties. 
The nonperturbative 
component that we find is consistent with a universal
quadratic (Gaussian) power-suppressed contribution of the kind that
may be expected on general grounds~\cite{Collins:2004nx}, and of a 
magnitude that is compatible with a previous global analysis 
of Drell-Yan $Q_T$ distributions~\cite{Konychev:2005iy}.

The D\O\ data are
precise enough and may be able to distinguish between the Gaussian 
and alternative nonperturbative functions that have been recently 
proposed~\cite{Collins:2013zsa}. It would be insightful to examine
constraints on a variety of the nonperturbative models that are
currently discussed~\cite{Aybat:2011zv,Becher:2011xn,Echevarria:2012pw,Schweitzer:2012hh}, 
as well as the $\sqrt{s}$ dependence of the
nonperturbative contributions by using a combination of the Tevatron
and LHC data. As such investigation demands significant computational
resources, it will be pursued in future work. 

Our main numerical results have been reported 
at the QCD Evolution Workshop at Thomas Jefferson National Accelerator 
Facility in May 2012~\cite{Guzzi:2012jc}. The current paper documents
this analysis in detail and  is organized as follows. 
Section~\ref{sec:Overview} reviews the relation between the
$\phi_\eta^*$ angle 
and transverse momentum $Q_T$ in the Collins-Soper-Sterman notations (Sec.~\ref{sec:phistar}),
general structure of the resummed cross section and 
estimation of NNLO contributions and their scale dependence 
(Secs.~\ref{gen-struc}, \ref{sec:WpertCanonical}, \ref{sec:WpertAS}), nonperturbative
model (Sec.~\ref{sec:WNP}), matching of the small-$Q_T$ and
large-$Q_T$ terms (Sec.~\ref{sec:Matching}), photon radiation contribution
(Sec.~\ref{sec:Photon}), and numerical accuracy (Sec.~\ref{sec:Num_acc}). 
In Sec.~\ref{sec:CSS_vs_BCFG}, distinctions between
the NNLL/NNLO resummed $Q_T$ distributions obtained in the CSS formalism 
and the alternative approach of Refs.~\cite{Bozzi:2005wk,Bozzi:2008bb,Catani:2012qa} are summarized.

Next, in Sec.~\ref{sec:numerical-results}, 
the size of the nonperturbative contributions is estimated by a $\chi^2$ analysis of the D\O\ data in three bins
of vector boson rapidity ($y_Z$), by applying two different methods to
examine the scale dependence of the resummed cross section.
By using the constraining power of this data set, we suggest a
Gaussian smearing factor suitable for $W$ and $Z$ production, and 
we give an estimate at 68\% C.L. for the leading parameter of the NP functional form. 
We provide the user with several sets of grids of theory 
predictions for phenomenological applications based on CT10 NNLO~\cite{Gao:2013xoa}
PDF eigenvector sets, and for scans of the nonperturbative smearing
function and estimates of its uncertainty in future measurements.

\section{Overview of the resummation method\label{sec:Overview}}

\subsection{Relation between $Q_T$ and $\phi^*_\eta$ variables \label{sec:phistar}}
The Collins-Soper-Sterman (CSS) resummation formalism predicts 
fully differential distributions in electroweak boson
production, including decay of heavy bosons. While the
original formulation of the CSS formalism deals with resummation of
logarithms dependent on the boson's transverse momentum $Q_T$, 
it can be readily extended to resum angular variables of
decay particles. One such variable is the azimuthal angle separation 
$\Delta\varphi$ of the leptons in the lab frame, which approaches $\pi$
(back-to-back production of leptons in the transverse plane) when
$Q_T\rightarrow 0$. Consequently, the region $\Delta\varphi \rightarrow
\pi$ is sensitive to small-$Q_T$ resummation~\cite{Balazs:1997xd}.  

Recently, an angular variable $\phi_\eta^*$ was proposed  
in~\cite{Banfi:2010cf} that has an experimental advantage 
compared to $Q_T$ and $\Delta\varphi$. The $\phi_\eta^*$ variable 
is not affected by the experimental resolution on the
magnitudes of the leptons' (transverse)  
momenta that limits the accuracy of the $Q_T$ measurement. Soft and
collinear resummation for the $\phi^*_\eta$ distribution can be worked
out either analytically~\cite{Banfi:2012du,Banfi:2011dx,Banfi:2011dm,Banfi:2009dy} or
numerically by integrating the resummed $Q_T$ distribution over the
leptons' phase space.

To describe decays of massive bosons,  
the CSS formalism~\cite{Balazs:1997xd}
usually operates with the lepton polar angle $\theta_{CS}$ 
and azimuthal angle $\varphi_{CS}$ 
in the Collins-Soper (CS) reference frame~\cite{Collins:1977iv}. The
CS frame is a rest frame of the vector boson in which the $z$
axis bisects the angle formed by the momenta $\vec p_1$ and -$\vec
p_2$ of the incident quark and antiquark. In the CS frame, 
the decay leptons escape back-to-back ($\vec l_1 + \vec l_2 = 0$), 
and the electron's and positron's 4-momenta are
\beq
  \left. l_1^\mu\right|_{\mbox{CS frame}} =(Q/2) \left\{ 1, \cos\varphi_{CS} \sin\theta_{CS}, \sin\varphi_{CS}
   \sin\theta_{CS}, \cos\theta_{CS} \right\}, 
\eeq
and
\beq
  \left. l_2^\mu\right|_{\mbox{CS frame}}=(Q/2) \left\{ 1, -\cos\varphi_{CS} \sin\theta_{CS}, -\sin\varphi_{CS}
   \sin\theta_{CS}, -\cos\theta_{CS}\right\}. 
\eeq

On the other hand, the angular variable $\phi_\eta^*$  is defined in
a different frame (``$\eta$ frame''), in which the leptons escape $\theta_\eta^*$ 
and $\pi-\theta_\eta^*$ with respect to the incident beams direction.
The $\eta$ frame is related to the lab frame by a boost $\beta=\tanh(\left(\eta_1 +\eta_2\right)/2)$ along 
the incident beam direction, where $\eta_1$ and $\eta_2$ are the pseudorapidities 
of $e^-$ and $e^+$ in the lab frame. The frame coincides with the CS frame when $Q_T=0$. 
Knowing the polar angle $\theta_\eta^*$ in the $\eta$ frame and the difference
$\Delta \varphi=\varphi_1-\varphi_2$ of the lepton's azimuthal angles 
in the transverse plane to the beam direction, one defines 
\beq
\phi^*_{\eta}=\tan{\left(\phi_{acop}/2\right)}\sin{\theta^*_{\eta}}
\label{def_phi}
\eeq
in terms of the acoplanarity angle $\phi_{acop} = \pi - \Delta \varphi$.
 We write $\cos\theta_\eta^*$ as a function of the lepton momenta in the lab
frame as
\beq
\cos\theta^*_{\eta} = \tanh\left( \frac{\eta_1 -
    \eta_2}{2}\right)=\frac{\sqrt{l^+_1 l_2^-} - 
\sqrt{l^-_1 l^+_2}}{\sqrt{l^+_1 l_2^-} + 
\sqrt{l^-_1 l^+_2}}=\frac{f\left(\cos\theta_{CS}\right) - f\left(-\cos\theta_{CS}\right)}{f\left(\cos\theta_{CS}\right) + f\left(-\cos\theta_{CS}\right)},
\label{sinthetaeta}
\eeq
where $l^\pm_{1,2} = (l^0_{1,2} \pm l^z_{1,2})/\sqrt{2}$,
\beq
f(\cos\theta_{CS})\equiv\sqrt{M_T^2+
2 M_T Q \cos \theta_{CS} +Q^2 \cos^2 \theta_{CS} 
-Q_T^2 \sin^2\theta_{CS} \cos^2\varphi _{CS}},
\label{f}
\eeq 
and $M_T^2=Q^2+Q_T^2$. We also write $\cos{\Delta \varphi}$ as
\begin{eqnarray}
&\cos{\Delta \varphi}=(Q_T^2-Q^2\sin^2\theta_{CS}-Q_T^2\sin^2\theta_{CS}\cos^2\varphi _{CS})\nonumber\\
&\times[(Q^2\sin^2\theta_{CS}+Q_T^2\sin^2\theta_{CS}\cos^2\varphi _{CS}+Q_T^2)^2-4M_T^2Q_T^2\sin^2\theta_{CS}\cos^2\varphi _{CS}]^{-\frac{1}{2}}.
\label{cosdeltaphi}
\end{eqnarray}

In the limit $Q_T \rightarrow 0$, $\phi^*_{\eta}$ simplifies to
\beq
\phi^*_{\eta} \approx (Q_T/Q)\sin\varphi_{CS},
\eeq
since $\tan(\phi_{acop}/2) =
\sqrt{\left(1+\cos\Delta\varphi\right)/\left(1-\cos\Delta\varphi\right)}$, and 
\bea
\theta^*_{\eta}\rightarrow \theta_{CS}, &&
\cos{\Delta \varphi}\rightarrow -1 + 2 \left(\frac{Q_T}{Q}\frac{\sin\varphi_{CS}}{\sin\theta_{CS}}\right)^2.
\eea
Measurement of $\phi^*_{\eta}$ thus directly probes
$Q_T/Q$.\footnote{The asymptotic
relation between $\phi^*_{\eta}$ and $Q_T/Q$ can alternatively be
obtained by introducing  the component $a_T$ 
of $\vec Q_T$ along the thrust axis 
$\hat{n}=(\vec{l}_{1,T}-\vec{l}_{2,T})/|\vec{l}_{1,T}-\vec{l}_{2,T}|$, where 
$\vec{l}_{1,T}$ and $\vec{l}_{2,T}$ are the transverse momenta of
$e^-$ and $e^+$, and 
identifying $a_T =Q_T\sin \varphi_{CS}$ at $Q_T \rightarrow 0$~\cite{Banfi:2011dx,Banfi:2009dy,Banfi:2010cf,Banfi:2008qs}.}

Relations like these can analytically express the 
$\phi^*_\eta$ distribution in terms of the $Q_T$ distribution, but in practice
it is easier to compute the $\phi^*_\eta$ distribution by
Monte-Carlo integration in \textsc{ResBos} code. In this case, the
interval of small $Q_T/Q$ maps onto the region of small $\phi^*_\eta$. 
For example, in $Z$ production at $Q\approx M_Z$,
the range $10^{-3}\leq \phi^*_\eta \leq 0.5$ radians corresponds 
to $0.1 \lesssim Q_T \lesssim 50$ GeV.

\subsection{General structure of the resummed cross section\label{gen-struc}}
The resummed cross sections that we present are based on the 
calculation in~\cite{Balazs:1995nz,Ladinsky:1993zn,Balazs:1997xd,Landry:2002ix} 
with added higher-order
radiative contributions (Secs.~\ref{sec:WpertCanonical},
\ref{sec:WpertAS}) and modified nonperturbative model (Sec.~\ref{sec:WNP}).
We write the fully differential cross section 
for $Z$ boson production and decay as
\ba
\frac{d\sigma\left(h_{1}h_{2}\rightarrow(Z\rightarrow\ell\bar{\ell})X\right)}{dQ^{2}\ dy_Z\ dQ_{T}^{2}\ d\cos\theta_{CS}\ d\varphi_{CS}}=
\sum_{\alpha=-1}^4 F_{\alpha}\left(Q,Q_T,y\right)A_{\alpha}\left(\theta_{CS},\varphi_{CS}\right)
\ea
in terms of the structure functions $F_{\alpha}(Q,Q_T,y_Z)$ and angular
functions $A_{\alpha}(\theta_{CS},\varphi_{CS})$. 
The variables $Q$, $Q_T$, and $y_Z$ correspond to
the invariant mass, transverse momentum, and rapidity of
the boson in the lab frame; $\theta_{CS}$ and
$\varphi_{CS}$ are the lepton decay angles in the CS frame. Among the 
structure functions $F_{\alpha}$, two 
(associated with the angular functions $A_{-1}=1+\cos^2\theta_{CS}$ and $A_3=2\cos\theta_{CS}$)
include resummation of soft 
and collinear logarithms in the small-$Q_T$ limit. For such functions, we write
\ba
F_{\alpha}(Q,Q_T,y_Z) = W_{\alpha}(Q, Q_T, y_Z; C_1/b, C_2 Q, C_3/b) +
Y_{\alpha}(Q, Q_T, y_Z; C_4 Q),
\ea
where 
\be
W_{\alpha}(Q, Q_T, y_Z) = \int\frac{d^{2}b}{4\pi^{2}}\, e^{i\vec{Q}_{T}\cdot\vec{b}}\sum_{j=u,d,s...}
\widetilde{W}_{\alpha,j}(b,Q,y_Z)
\label{resummed1}
\ee
is introduced to resum small-$Q_T$ logarithms to all orders in $\alpha_s$.  
The $W$ term depends on several auxiliary QCD scales $C_1/b$, $C_2 Q$,
and $C_3/b$ with constant coefficients $C_{1,2,3}\approx 1$ that 
emerge from the solution of differential equations describing
renormalization and gauge invariance of $Q_T$ distributions~\cite{Collins:1981uk,CollinsBook2}.
$Y_{\alpha}(Q, Q_T, y_Z; C_4 Q)$ is a part of the non-singular remainder, 
or ``the $Y$ term'''. It depends on a factorization and
renormalization momentum scale $C_4 Q$. 

The Fourier-Bessel integral over
the transverse position $b$ in the $W$ term in Eq.~(\ref{resummed1})
acquires contributions from the region of small transverse positions
$0 \leq b \lesssim 1\mbox{ GeV}^{-1}$, where the form factor 
can be approximated in perturbative QCD, and the
region $b \gtrsim 1\mbox{ GeV}^{-1}$, where the perturbative
expansion in the QCD coupling $\alpha_s(1/b)$ breaks down, and
nonperturbative methods are necessitated. 
In $Z$ boson production, the small-$b$ perturbative contribution 
dominates the Fourier-Bessel integral for any $Q_T$ value~\cite{Curci:1979bg,Konychev:2005iy,Arnold:1990yk}.
At $Q_T$ below 5 GeV, the production rate is also mildly 
sensitive to the behavior in the $b> 0.5$ GeV$^{-1}$ interval, where
the full expression for $\widetilde{W}_{\alpha,j}(b,Q)$ is
yet unknown. 

To determine the acceptable large-$b$ forms
of $\widetilde{W}_{\alpha,j}(b,Q)$ by comparison to the latest $Z$ boson
data, we need to update the leading-power contribution to
$\widetilde{W}_{\alpha,j}(b,Q,y_Z)$ computable in perturbative QCD,
denoted by $\widetilde{W}_{\alpha,j}^{pert}(b,Q,y_Z)$, 
by considering
additional QCD and electromagnetic corrections and dependence
on QCD factorization scales. In particular, 
scale dependence in the perturbative form factor
$\widetilde{W}^{pert}$ may smear the sensitivity to the nonperturbative 
factor~\cite{Catani:2009sm,Banfi:2012du,Catani:2012qa,Banfi:2011dm}. 
We will review the perturbative contributions in the next two subsections. 

\subsection{Perturbative coefficients for canonical scales \label{sec:WpertCanonical}}
For a particular ``canonical'' combination
of the scale parameters, the perturbative contributions simplify; 
the resummed form factor at $b\ll 1 \mbox{ GeV}^{-1}$ 
takes the form 
\begin{eqnarray}
\widetilde{W}^{pert}_{\alpha,j}(b,Q,y_Z)& = & \sum_{j=u,d,s...}\left|H_{\alpha,j}(Q,\Omega)\right|^{2}\,
\exp\left[-S(b,Q)\right]
\nonumber \\
 & \times & \sum_{a=g,q,\bar{q}}\left[{\cal C}_{ja}\otimes 
f_{a/h_{1}}\right] \left(\chi_{1},\mu_F\right)
\sum_{b=g,q,\bar{q}}\left[{\cal C}_{\bar{j}b}\otimes f_{b/h_{2}}\right]
\left(\chi_{2},\mu_F\right)
\label{defWc}
\end{eqnarray}
in terms of a   
$2\rightarrow 2$ 
hard part $\vert H_{\alpha,j}(Q,\Omega)\vert^2$, Sudakov integral
\be
S(b,Q)=\int_{b_0^2/b^2}^{Q^2}
\frac{d\bar{\mu}^{2}}{\bar{\mu}^{2}}\left[A(\bar{\mu})\,\ln\left(\frac{Q^2}{\bar{\mu}^{2}}\right)+B(\bar{\mu})\right],
\label{Sudakov}
\ee
and convolutions $\left[{\cal C}_{j/a}\otimes f_{a/h}\right]$ of
Wilson coefficient functions ${\cal C}_{j/a}$
and PDFs $f_{a/h}$ for a parton
$a$ inside the initial-state hadron $h$.  
The convolution integral is defined by 
\ba
&&\left[{\cal C}_{j a}\otimes f_{a/h}\right](\chi,\mu_F)=\int_{x}^{1}\frac{d\xi}{\xi}
{\cal C}_{j a}\left(\frac{\chi}{\xi},\mu_F\right)f_{a/h}(\xi,\mu_F).
\label{conv}
\ea
In Eq.~(\ref{conv}) the convolution depends on the momentum
fractions $\chi_{1,2}$ that reduce to $x_{1,2}^{(0)}\equiv
(Q/\sqrt{s}) e^{\pm y_Z}$ in the limit $Q_T^2/Q^2 \rightarrow 0$, as explained in Sec.~\ref{sec:Matching}, as well as on the factorization
scale $\mu_F=b_0/b$. Some scales are proportional to the constant $b_0=2e^{-\gamma_E}=1.123...$, where $\gamma_E= 0.577...$ is the Euler-Mascheroni constant. 

The functions $H_{\alpha,j}$, $A$, $B$, and $C$ can be expanded as a series in
the QCD coupling strength,
\be
H_{\alpha,j}(Q,\Omega; \alpha_s(\bar{\mu})) =
1+ \sum_{n=1}^\infty\left(\frac{\alpha_s(\bar{\mu})}{\pi}\right)^n H^{(n)}_{\alpha,j}(Q,\Omega)\, ,\quad\quad
A(\alpha_s(\bar{\mu})) =
\sum_{n=1}^\infty\left(\frac{\alpha_s(\bar{\mu})}{\pi}\right)^n A^{(n)}\,, \mbox{etc.}
\ee
Some perturbative contributions can be moved between the hard function $H_{\alpha,j}$ 
and Sudakov exponential depending on the resummation scheme~\cite{Catani:2000vq}. 
In the Collins-Soper-Sterman (CSS) resummation scheme,
$H_{\alpha,j}(\alpha_s)=1$ to all $\alpha_s$ orders. In the Catani-De
Florian-Grazzini (CFG) resummation scheme, $H_{\alpha,j}(\alpha_s)$ includes
hard virtual contributions starting at ${\cal O}(\alpha_s)$, while 
the Sudakov exponential depends only on the type of the
initial-state particle (quark or gluon) that radiates soft emissions. 
In Drell-Yan production, differences between the CSS and CFG schemes 
are small, below 1\% in the kinematic region explored.
We carry out the analysis in the CSS scheme, but 
the nonperturbative function that we obtain can be readily 
used with the CFG scheme.

The functions $A$ and $B$ for the canonical choice of scales are evaluated up 
to ${\cal O}(\alpha_s^3)$ and ${\cal O}(\alpha_s^2)$ respectively, 
using their known perturbative coefficients~\cite{Kodaira:1981nh,Kodaira:1982az,
Davies:1984hs,Davies:1984sp,Catani:1988vd,Moch:2004pa}. 
The three-loop coefficient $A^{(3)}$ 
is included, but has a weak effect on the cross section 
(3\% at $Q_T\approx 2$ GeV). The coefficient $A^{(3)}$ has 
been derived within the soft-collinear effective theory~\cite{Becher:2010tm} and found to contain a 
term arising from the ``collinear anomaly'', besides 
the ${\cal O}(\alpha_s^3)$ cusp contribution known from~\cite{Moch:2004pa}.
The ``collinear'' anomaly contribution breaks the symmetry of the SCET 
Lagrangian by regulators of loop 
integrals~\cite{Becher:2010tm,Becher:2006mr,Becher:2006nr,Becher:2007ty}. 
The expansion of $\widetilde W^{pert}$ in the CSS scheme, 
is found to be in agreement with that derived in SCET up to NLO. 
We checked that $A^{(3)}_{SCET}$ has inappreciable influence 
on the conclusions.

The Wilson coefficient functions ${\cal C}^{(i)}$ are computed 
exactly up to ${{\cal O}(\alpha_s)}$ and approximately 
to ${\cal O}(\alpha_s^2)$. Most of our numerical results were obtained with the
${\cal O}(\alpha_s^2)$ approximation for the Wilson coefficient
before the exact ${\cal O}(\alpha_s^2)$ result 
were published~\cite{Bozzi:2010xn, Catani:2009sm, Catani:2012qa}.
This expression is constructed by using a numerical approximation for
the canonical part of the Wilson coefficient at ${\cal O}(\alpha_s^2)$ and exact
expression for its dependence on soft scales. 
Our {\it a posteriori} comparison shows the approximation to be close
to the exact expression, cf. the
next subsection.

The $Y$ contribution in Eq.~(\ref{resummed1}) is defined as 
the difference between the fixed-order perturbative $Q_T$ 
distribution calculation and the asymptotic distribution
obtained by expanding the perturbative part $\widetilde{W}^{pert}$ up to the same order. It is given by
\ba
&& Y_\alpha(Q_T,Q,y_Z)=\int \frac{d\xi_1}{\xi_1}\int \frac{d\xi_2}{\xi_2}
\sum_{n=1}^{\infty}\left[\frac{\alpha_s(C_4 Q)}{\pi}\right]^n\times
\nonumber\\
&& f_{a/h_1}(\xi_1, C_4 Q) ~R_{\alpha,ab}^{(n)}\left(Q_T,Q,y_Z; \xi_1,
\xi_2, C_4 Q \right)~
f_{b/h_2}(\xi_2,C_4 Q),
\ea
where the functions $R_{\alpha,ab}^{(n)}$ are integrable when
$Q_T\rightarrow 0$, and their explicit expressions for all
contributing $\alpha$ to ${\cal O}(\alpha_s)$
can be found in~\cite{Collins:1984kg, Balazs:1997xd}.
The $O(\alpha_s^2)$ contribution to the dominant structure function 
$Y_{-1}$ is included using the calculation 
in~\cite{Arnold:1988dp, Arnold:1990yk}. $O(\alpha_s^2)$ corrections 
to the other structure functions in the $Y$ term 
are essentially negligible in the small-$Q_T$ region of our fit.

\subsection{Perturbative coefficients for arbitrary scales\label{sec:WpertAS}}

The resummed form factor in Eq.~(\ref{defWc}) can be
generalized to allow variations in the arbitrary factorization scales
arising in the solution of Collins-Soper differential equations. At
small $b$, the scale-dependent expression takes the form 
\begin{eqnarray}
\widetilde{W}^{pert}_{\alpha,j} & = & \sum_{j=u,d,s...}\left|H_{\alpha,j}(Q,\Omega,C_2 Q)\right|^{2}\,
\exp\left[-\int_{C_{1}^{2}/b^{2}}^{C_{2}^{2}Q^{2}}
\frac{d\bar{\mu}^{2}}{\bar{\mu}^{2}}A(\bar{\mu};C_{1})\,\ln\left(\frac{C_{2}^{2}Q^{2}}{\bar{\mu}^{2}}\right)+B(\bar{\mu};C_{1},C_{2})\right]
\nonumber \\
 & \times & \sum_{a=g,q,\bar{q}}\left[{\cal C}_{ja}\otimes 
f_{a/h_{1}}\right]\left(\chi_{1},\frac{C_{1}}{C_{2}},\frac{C_{3}}{b}\right)
\sum_{b=g,q,\bar{q}}\left[{\cal C}_{\bar{j}b}\otimes f_{b/h_{2}}\right]
\left(\chi_{2},\frac{C_{1}}{C_{2}},\frac{C_{3}}{b}\right),
\label{defW}
\end{eqnarray}
where the coefficients
$C_1=b \bar{\mu}$ and $C_2=\bar{\mu}/Q$ are associated with 
the lower and upper integration limits in Eq.~(\ref{defW}), 
while $\mu_F=C_3/b$ is the factorization scale at which Wilson coefficient functions 
are evaluated. The ``canonical'' representation 
adopted in Eq.~(\ref{defWc}) corresponds to $C_1=C_3=b_0$ and $C_2=1$.
For the rest of the discussion, we use the same scale $C_2 Q$ to
compute the hard function $H_{\alpha,j}$ and the $Y$ term, {\it i.e.}
set $C_4=C_2$.

The perturbative coefficients $A^{(n)}$, $B^{(n)}$, and $C^{(n)}$ 
are generally dependent on the scale coefficients, but the full form
factor $\widetilde{W}^{pert}$ is independent when expanded to a fixed
order in $\alpha_s$. We can therefore reconstruct the perturbative
coefficients order-by-order for arbitrary $C_1$, $C_2$, $C_3$ if we
know the canonical values of the coefficients, indicated by the
superscript ``(c)''.

By truncating the series at ${\cal O}(\alpha_s^2)$, we must have 
\beq
\widetilde{W}(b,Q,C_1,C_2,C_3)\vert_{{\cal O}(\alpha_s^2)}=\widetilde{W}(b,Q,C_1=C_3=b_0,C_2=1)\vert_{{\cal O}(\alpha_s^2)}.
\label{WeWc}
\eeq
Making a series expansion on both sides of Eq.~(\ref{WeWc}),
we find the following relations by equating the coefficients in
front of each power of $ \log\left(b^2 Q^2\right)$:
\begin{eqnarray}
 A^{(1)}(C_{1}) &=&  A^{(1,c)};\label{A1A1c}\\
A^{(2)}(C_{1}) &=&  A^{(2,c)}-A^{(1,c)}\beta_{0}\ln\frac{b_0}{C_{1}};\\
A^{(3)}(C_{1}) &=&  A^{(3,c)}-2A^{(2,c)}\beta_{0}\ln\frac{b_0}{C_{1}}-\frac{A^{(1,c)}}{2}\beta_{1}\ln\frac{b_0}{C_{1}}+A^{(1,c)}\beta_{0}^{2}\left(\ln\frac{b_0}{C_{1}}\right)^{2};\\
B^{(1)}(C_{1},C_{2}) &=& B^{(1,c)}-A^{(1,c)}\ln\frac{b_0^{2}C_{2}^{2}}{C_{1}^{2}};\\
B^{(2)}(C_{1},C_{2}) &=&  B^{(2,c)}-A^{(2,c)}\ln\frac{b_0^{2}C_{2}^{2}}{C_{1}^{2}} \nonumber \\
&+&  \beta_{0}\left[A^{(1,c)}\ln^{2}\frac{b_0}{C_{1}}+B^{(1,c)}\ln C_{2}-A^{(1,c)}\ln^{2}C_{2}\right];
\end{eqnarray}
\begin{eqnarray}
\mathcal{C}_{ja}^{(1)}\left(\xi,\frac{C_{1}}{C_{2}},C_3\right) &=&  
\mathcal{C}_{ja}^{(1,c)}(\xi)
+\delta_{ja}\delta(1-\xi)\, \left\{\frac{B^{(1,c)}}{2}\ln\frac{b_0^{2}C_{2}^{2}}{C_{1}^{2}}
-\frac{A^{(1,c)}}{4}\left(\ln\frac{b_0^{2}C_{2}^{2}}{C_{1}^{2}}\right)^{2}
\right\} \nonumber \\ 
& -& P_{ja}^{(1)}(x)\ln\frac{\mu_F b}{b_0}; \\
\mathcal{C}_{ja}^{(2)}\left(\xi,\frac{C_{1}}{C_{2}},C_{3}\right) &=&  \mathcal{C}_{ja}^{(2,c)}(\xi) 
 + \delta_{ja}\delta(1-\xi) L^{(2)}(C_1,C_2)\nonumber \\
 & + & \left\{ \frac{\beta_{0}}{2}{\cal C}_{jb}^{(1,c)}(\xi)-[{\cal C}_{jb}^{(1,c)}\otimes P_{ba}^{(1)}](\xi)-P_{ja}^{(2)}(\xi)\right\} \ln\frac{\mu_F b}{b_{0}}\nonumber\\
&+& \frac{1}{2}[P_{jb}^{(1)}\otimes P_{ba}^{(1)}](\xi)\,\ln^{2}\frac{\mu_{F}b}{b_{0}}.\label{WC2}
\end{eqnarray}
Here the beta-function coefficients for $N_c$ colors and $N_f$
flavors are $\beta_{0}=(11N_{c}-2N_{f})/6$,
$\beta_{1}=(17N_{c}^{2}-5N_{c}N_{f}-3C_{F}N_{f})/6$,
$C_{F}=(N_{c}^{2}-1)/(2N_{c})$. $P_{ja}^{(n)}(\xi)$ is a splitting
function of order $n$. 
The term $L^{(2)}(C_1,C_2)$  in $\mathcal{C}_{ja}^{(2)}$ 
realizes the exact dependence on the {\it soft} scale constants $C_1$
and $C_2$:
\bea
&&L^{(2)}(C_1,C_2) \equiv \frac{1}{32}(A^{(1,c)})^{2}\log^{4}\left(\frac{b_{0}^{2}C_{2}^{2}}{C_{1}^{2}}\right) \nonumber\\
&-&\frac{1}{8}A^{(1,c)}\beta_{0}\log\left(\frac{b^{2}\mu_{F}^{2}}{b_{0}^{2}}\right)\log^{2}\left(\frac{b_{0}^{2}C_{2}^{2}}{C_{1}^{2}}\right)
-  \frac{1}{8}A^{(1,c)}B^{(1,c)}\log^{3}\left(\frac{b_{0}^{2}C_{2}^{2}}{C_{1}^{2}}\right) \nonumber\\
&-&\frac{1}{24}A^{(1,c)}\beta_{0}\log^{3}\left(\frac{b_{0}^{2}C_{2}^{2}}{C_{1}^{2}}\right)-\frac{1}{4}A^{(1,c)}\delta{\cal C}_{1c}\log^{2}\left(\frac{b_{0}^{2}C_{2}^{2}}{C_{1}^{2}}\right)\nonumber \\
 &-&  \frac{1}{4}A^{(2,c)}\log^{2}\left(\frac{b_{0}^{2}C_{2}^{2}}{C_{1}^{2}}\right)+\frac{1}{4}\beta_{0}B^{(1,c)}\log\left(\frac{b^{2}\mu_{F}^{2}}{b_{0}^{2}}\right)\log\left(\frac{b_{0}^{2}C_{2}^{2}}{C_{1}^{2}}\right)\nonumber\\
&+&\frac{1}{8}(B^{(1,c)})^{2}\log^{2}\left(\frac{b_{0}^{2}C_{2}^{2}}{C_{1}^{2}}\right)
 +  \frac{1}{8}\beta_{0}B^{(1,c)}\log^{2}\left(\frac{b_{0}^{2}C_{2}^{2}}{C_{1}^{2}}\right)\nonumber \\
&+&\frac{1}{2}B^{(1,c)}\delta{\cal C}_{1c}\log\left(\frac{b_{0}^{2}C_{2}^{2}}{C_{1}^{2}}\right)+\frac{1}{2}B^{(2,c)}\log\left(\frac{b_{0}^{2}C_{2}^{2}}{C_{1}^{2}}\right).
\label{LC1C2}
\eea

The dependence on $C_3$ is small already at ${\cal O}(\alpha_s)$.
The canonical coefficients in the CSS scheme are~\cite{Balazs:1997xd}
\ba
&&A^{(1,c)} = C_F; \hspace{1cm} B^{(1,c)}=-\frac{3}{2} C_F;
\hspace{1cm} A^{(2,c)} = C_F\left[\left(\frac{67}{36}-\frac{\pi^2}{12}\right) C_A -\frac{5}{18}N_f\right];
\nonumber\\
&&B^{(2,c)}=C_F^2 \left( \frac{\pi^2}{4} -\frac{3}{16} -3 \zeta_3 \right)
+C_A C_F\left( \frac{11}{36}\pi^2 - \frac{193}{48} + \frac{3}{2}\zeta_3 \right)
+\frac{1}{2} C_F N_f \left( -\frac{\pi^2}{9} + \frac{17}{12}\right),
\nonumber\\
\ea
and $\delta\mathcal{C}^{(1,c)}=-\ln^2(C_1/(b_0 C_2)e^{-3/4}) + \pi^2/4 - 23/16$.

The expression for
$\mathcal{C}_{ja}^{(2)}\left(\xi,C_{1}/C_{2},C_{3}\right)$ in
Eq.~(\ref{WC2}) is more complex than the one for the other
coefficients. From the fixed-order NNLO calculation~\cite{Melnikov:2006kv} we know that the contribution ${{\cal C}}_{ja}^{(2)}$
is small in magnitude (2-3\% of the cross section) in $Z$ production
and does not vary strongly with $y_Z$~\cite{Anastasiou:2003ds},
hence has weak dependence on $\xi$. Its importance is further reduced
in the computation of the {\it normalized} $\phi_\eta^*$ distributions
that we will work with.

Knowing this, we approximate
$\mathcal{C}_{ja}^{(2)}\left(\xi,C_{1}/C_{2},C_{3}\right)$ as
\begin{equation}
{{\cal C}}_{ja}^{(2)}(\xi,C_{1}/C_{2},C_{3}) \approx \Biggl\{
\langle \delta {\cal C}^{(2,c)}\rangle
+L^{(2)}(C_1,C_2)
\Biggr\}\, \delta(1-\xi)\,\delta_{ja},
\label{C2approx}
\end{equation}
where $\langle \delta {\cal C}^{(2,c)}\rangle$ denotes the average value of
the Wilson coefficient in $Z$ production for the canonical scale
combination and $L^{(2)}(C_1,C_2)$ is the same as in Eq.(\ref{LC1C2}). 
It is estimated from the requirement that the resummed
cross section reproduces the fixed-order prediction for
the computation of the invariant mass distribution, which 
is known since a long time~\cite{Hamberg:1990np} 
and was evaluated in our analysis by the computer code \textsc{Candia}~\cite{Cafarella:2007tj,Cafarella:2008du} 
\footnote{Other computer codes are also publicly available at this purpose: 
\textsc{DYNNLO}~\cite{Catani:2009sm,Catani:2007vq} 
and \textsc{Vrap}~\cite{Anastasiou:2003ds}.}.
The second term in Eq.~(\ref{C2approx}) realizes the exact dependence on {\it soft} scale constants $C_1$
and $C_2$. The $\xi$ dependence of ${{\cal C}}_{ja}^{(2)}(\xi,C_{1}/C_{2},C_{3})$ is neglected in this
approximation. The $C_3$ dependence is included to ${\cal O}(\alpha_s)$ 
and is of the same order as the ${\cal O}(\alpha_s^2)$
dependence on $C_1$ and $C_2$. 

\begin{figure}[tb]
\begin{center}
\includegraphics[width=13cm, angle=0]{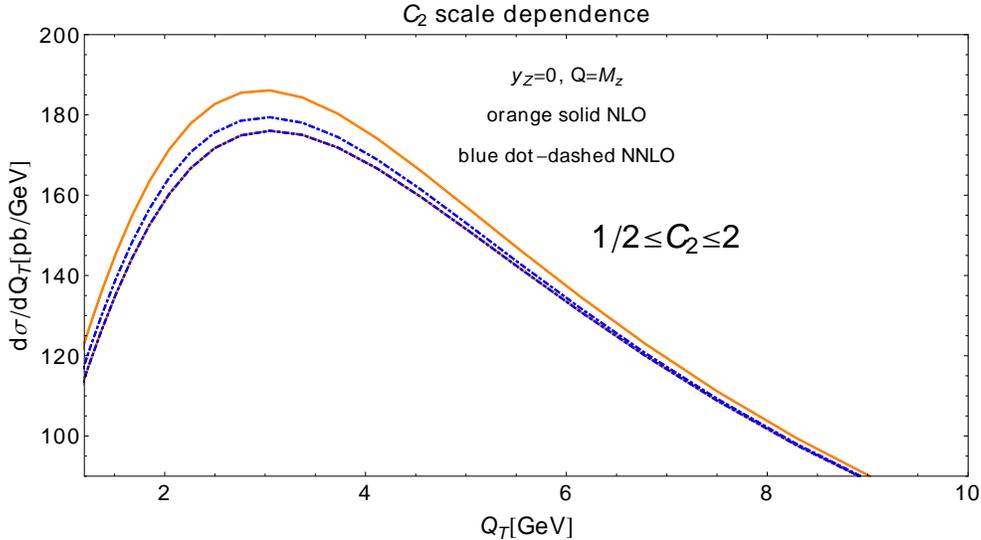}
\caption{Dependence of $Z$ boson $Q_T$ distribution on the
    scale parameter $C_2$ at ${\cal O}(\alpha_s^2)$ and ${\cal O}(\alpha_s)$.
\label{fig:C2dep}}
\end{center}
\end{figure}

The part $\delta {\cal C}^{(2,c)}_{ja}$ of ${\cal C}^{(2,c)}_{ja}$ proportional to
$\delta_{ja}\delta(1-\xi)$ can be determined from the calculation
in~\cite{Catani:2012qa} as
\ba
\delta{\cal C}^{(2)}_{qq,c}&=&
C_A C_F \left(\frac{59 }{18}\zeta_3-\frac{1535}{192}+\frac{215 }{216}\pi ^2-\frac{\pi ^4}{240}\right)+\frac{1}{4} C_F^2
\left(-15 \zeta_3+\frac{511}{16}-\frac{67 \pi ^2}{12}+\frac{17 }{45}\pi ^4\right)
\nonumber\\
&-&\frac{1}{16} \left(\pi ^2-8\right)^2 C_F^2+\frac{1}{864} C_F N_f \left(192 \zeta_3+1143-152 \pi ^2\right),
\ea
where $\zeta_3=1.20206...$, $C_F=(N_C^2-1)/(2N_C)$, $C_A=N_C$. Using the following relation in the CFG scheme,
\ba
H_{\alpha,j}^{DY}=1+\frac{\alpha_s(Q)}{\pi} H^{DY(1)}
+\frac{\alpha_s^2(Q)}{\pi^2} H^{DY(2)} + \dots,
\ea
one can estimate that the impact on $H^{DY}_q$ due to the inclusion of the ${\cal O}(\alpha_s^2)$ 
virtual corrections $H^{DY(2)}_q$ at $Q \approx M_Z$ is about
2\%. This correction is of the same order as the magnitude of
the effect of about 1\% from the averaged coefficient  $\langle \delta {\cal C}^{(2,c)}\rangle$ in
our calculation. This approximation is valid in the 
kinematic region of $W/Z$ production.
The full expression for ${\cal C}^{(2,c)}_{ja}(\xi)$
can be implemented in the future numerical work when the experimental
errors further decrease. 

The effect of the inclusion of scale-dependent terms at ${\cal O}(\alpha_s^2)$
is illustrated in Fig.~\ref{fig:C2dep} for the $Q_T$ differential cross section 
for Tevatron $Z$ production at the central rapidity $y_Z=0$ and $Q=M_Z$.
The orange solid band is the ${\cal O}(\alpha_s)$ uncertainty obtained
by variations of $C_2$ in the range $0.5-2$, while
the blue dot-dashed band is the same uncertainty evaluated at 
${\cal O}(\alpha_s^2)$. The sensitivity of the cross section 
to $C_2$ is clearly reduced upon the inclusion of the ${\cal
  O}(\alpha_s^2)$ contribution.  

\subsection{Nonperturbative resummed contributions \label{sec:WNP}}
Our fit to the $\phi^*_\eta$ will adopt a simple flexible
convention~\cite{Konychev:2005iy} for $\widetilde W_\alpha(b,Q)$ at
$b\gtrsim 1\mbox{ GeV}^{-1}$ that can emulate a variety of
functional forms arising in detailed nonperturbative 
models~\cite{Korchemsky:1994is,Ellis:1997ii,Guffanti:2000ep,Qiu:2000hf,Tafat:2001in,Kulesza:2002rh,Becher:2011xn,GarciaEchevarria:2011rb,Echevarria:2012pw}.

The convention is motivated by the
observation that, given the strong suppression of the deeply 
nonperturbative large-$b$ region in $Z$ boson production, only contributions 
from the transition region of $b$ of about $1$ GeV$^{-1}$
are non-negligible compared to the perturbative contribution 
from $b< 1\mbox{ GeV}^{-1}$. 
In the transition region, $\widetilde W(b,Q)$ can be 
reasonably approximated by the extrapolated leading-power, or perturbative, 
part $\widetilde{W}^{pert}(b,Q)$, and the nonperturbative
smearing factor $\widetilde{W}^{NP}(b,Q)$:
\be
\widetilde{W}_{\alpha,j}(b,Q,y_Z)=\widetilde{W}_{\alpha,j}^{pert}(b_*,Q,y_Z)\widetilde{W}^{NP}(b,Q,y_Z).
\label{WpertWNP}
\ee
When $b$ is large, 
the slow $b$ dependence in $\widetilde{W}_{\alpha,j}^{pert}(b_*,Q)$
can be neglected, compared to the rapidly changing
$\widetilde{W}_{NP}(b,Q)$. The latter contribution 
captures the effect of the powerlike contributions
proportional to $b^p$ with $p>0$ 
that alter the large-$b$ tail 
of $\widetilde{W}(b,Q)$ in a different way compared to
 $\widetilde{W}^{pert}(b,Q)$. The powerlike contributions suppress the rate only at $Q_T$ below
2-3 GeV, while the leading-power term and its scale
dependence affect a broader interval of $Q_T$ values 
(see representative figures in Ref.~\cite{Banfi:2011dm}). 
The nonperturbative suppression results in a characteristic shift of the peak in the $d\sigma/dQ_T$ 
distribution, which is distinct from the scale dependence.

To avoid divergence due to the Landau pole in
$\alpha_s(\overline\mu)$ at $\overline\mu \rightarrow 0$, we redefine
the scales of order $1/b$ in $\widetilde W^{pert}(b,Q)$ according to 
the $b_*$ prescription~\cite{Collins:1981va,Collins:1984kg}
dependent on two parameters~\cite{Konychev:2005iy}. 
In the Sudakov exponential, the lower limit 
$(C_1/b)^2$ is replaced by $\left(C_1/b_*(b,b_{max})\right)^2$, with
\be
b_*(b,b_{max}) \equiv \frac{b}{\sqrt{1+(b/b_{max})^2}},
\ee
where $b_{max}$ is set to $1.5\mbox{ GeV}^{-1}$ in~\cite{Konychev:2005iy}.
To avoid evaluating the PDFs $f_{a/h}(\xi,\mu_F)$ at a factorization
scale $\mu_F$ below the initial PDF scale $\mu_{ini}\approx 1\mbox{
  GeV}$, we choose $\mu_F=C_3/b_*(b,C_3/\mu_{ini})$; it
is larger than $\mu_{ini}$ for any $b$.
This prescription is preferred by the global fit to Drell-Yan
$Q_{T}$ data, where it both 
preserves the exact perturbative expansion for $\widetilde{W}^{pert}$
at $b<1$ GeV$^{-1}$ and improves the agreement with the data.

In a broad range of $Q$ values in the Drell-Yan process, 
the behavior of experimentally observed $Q_T$ distributions is
described by~\cite{Landry:2002ix,Konychev:2005iy}
\be
\widetilde{W}^{NP}\left(b,Q\right)=\textrm{exp}
\left[-b^2\left(a_{1} + a_{2}  \ln\left(\frac{Q}{2\ Q_0}\right) 
+ a_{3}  \ln\left(\frac{x^{(0)}_1 x^{(0)}_2}{0.01}\right)\right)\right]\,,
\label{WNPKN}
\ee
with $x^{(0)}_{1,2}=\frac{Q}{\sqrt{s}}e^{\pm y}$, free parameters $a_1,\
a_2,\ a_3$, and a fixed dimensional parameter $Q_0=1.6$ GeV.
The $b^2$ dependence characterizes the leading power-suppressed
contribution~\cite{Korchemsky:1994is} that can be resolved with the
available data. The $\ln(Q)$ dependence
is predicted by the Collins-Soper evolution equation~\cite{Collins:1981uk}. 
The higher-order power-suppressed contributions 
proportional to $b^4$, etc. cannot be reliably distinguished 
in the fit from the $b^2$ term. Although linear contributions
proportional to $b$ may also arise from long-distance dynamics~\cite{Schweitzer:2012hh}, 
they have been empirically disfavored in a
global $Q_T$ fit~\cite{Landry:2002ix}.

In the vicinity of $Q$ around $M_{Z}$, Eq.~(\ref{WNPKN})
reduces to 
\be
\widetilde{W}^{NP}\left(b,Q\approx M_{Z}\right)=\textrm{exp}\left[-b^2
  a_Z\right]
\ee
with
\be
a_{Z} = a_{1} + a_{2}  \ln\left(\frac{M_Z}{2\ Q_0}\right) 
+ a_{3}  \ln\left(\frac{M_Z^2}{0.01~s}\right).
\label{WNPGNW}
\ee

One of the essential applications of CSS resummation formalism
concerns the measurement of $W$
boson mass in hadron-hadron collisions. 
The current most precise $W$ mass measurements obtained by the D\O\ and CDF collaborations 
at the Tevatron~\cite{Abazov:2012bv,Aaltonen:2012bp} quote a total
error of about 20 MeV, with the bulk of it (approximately 90\%)
associated with three theoretical sources: PDF uncertainty (of order
10 MeV according to~\cite{Bozzi:2011ww}), EW corrections, and the
model of $\widetilde{W}_{NP}(b,Q)$ in production of $W$ bosons.
The last source of uncertainty appears because
the $W$ mass measurements are sensitive to the shape of the
cross section in the low-$Q_T$ region.  

Once $a_Z$ is determined from $Z/\gamma^*$ boson production, 
it is easy to predict $\widetilde{W}^{NP}$ in $W$ boson production at the same $\sqrt{s}$: 
\be
\widetilde{W}^{NP}\left(b,Q\approx M_{W}\right)=\textrm{exp}\left[-b^2 a_W\right],
\ee
where
\begin{equation}
a_{W} = a_Z  + a_{2}  \ln\left(\frac{M_W}{M_Z}\right) + a_{3}  \ln\left(\frac{M_W^2}{M_Z^2}\right).
\label{WNPW}
\end{equation}
For $b_{max}=1.5$ GeV$^{-1}$, one finds $a_{2}=0.17\pm 0.03$ GeV$^{2}$
and $a_3=-0.03\pm 0.02$ GeV$^{2}$~\cite{Konychev:2005iy}, where the
error estimate includes the scale dependence. The log terms
 proportional to $a_2$ and $a_3$ are small in Eq.~(\ref{WNPW}),
so that it is safe to assume $a_W\approx a_Z$ in central-rapidity 
measurements at the same $\sqrt{s}$.

If $Q$ is substantially different from $M_Z$, or if predictions for the
LHC are made, the $a_2$ and $a_3$ contributions cannot be
neglected. The nonperturbative coefficient becomes 
\begin{equation}
a(Q,\sqrt{s}) = a_Z(1.96\mbox{ TeV})  + a_{2}  \ln\left(\frac{Q}{M_Z}\right) + a_{3}
\ln\left(\frac{Q^2}{M_Z^2}\frac{s}{(1.96 \mbox{ TeV})^2}\right).
\label{WNPQ}
\end{equation}

\subsection{Matching the $W$ and $Y$ terms \label{sec:Matching}}

By examining the mapping of $Q_T$ distributions on $\phi^*_\eta$
distributions discussed in Sec.~\ref{sec:phistar}, 
we can identify three regions with
distinct QCD dynamics:  the resummation region $\phi^*_\eta \lesssim
0.1$ rad, where the $W$ term dominates; the intermediate (matching)
region $0.1 \lesssim \phi^*_\eta \lesssim 0.5$ rad; and the perturbative
region $\phi^*_\eta \gtrsim 0.5$ rad, where the $W+Y$ term 
approaches the fixed-order (FO) contribution. 
As $\phi^*_\eta$ increases in the intermediate region, 
the $W+Y$ term eventually becomes smaller than the FO term at 
$\phi^*_\eta\equiv\phi_{switch}(Q,y_Z)$. The final cross
section is taken to be equal to the $W+Y$ term at 
$\phi^*_\eta < \phi_{switch}$ and FO term at $\phi^*_\eta \geq
\phi_{switch}$~\cite{Balazs:1997xd}. 

The position of the switching point is subject to some variations 
dependent on the shapes of the $W$ term and its asymptotic expansion at
not too small $\phi^*_\eta \propto Q_T/Q$, {\it i.e.} away from the
$Q_T^2/Q^2 \rightarrow 0$ limit where the $W$ term is uniquely
defined. These variations have almost no effect on the fit of the
nonperturbative function in the resummation region $\phi_\eta^*$. 
They originate from the possibility of including additional
terms of order $Q_T^2/Q^2$ in the longitudinal momentum fractions
$\chi_{1,2}$ in the $W$ term and its asymptotic expansion. These terms 
vanish at $Q_T^2/Q^2\rightarrow 0$, but
they can be numerically important or even desirable in the
intermediate region, where they may improve agreement between the $W+Y$ and
FO terms. 

At intermediate $Q_T/Q$, radiation of a $Z$
boson and semi-hard jets requires sufficient center-of-mass 
energy of incident partons, 
or large enough partonic momentum fractions $\xi_1$ and
$\xi_2$. For example, the FO hadronic cross
section is written as 
\ba
&&\frac{d\sigma}{dQ^2 dy_Z dQ_T^2}=\sum_{a,b}\int_0^1 d\xi_1 \int_0^1
d\xi_2\frac{d\hat{\sigma}}{dQ^2dy_Z dQ_T^2}f_{a/A}(\xi_1)
f_{b/B}(\xi_2) 
\nonumber\\
&&\equiv\int_{\bar \xi_1}^1 d\xi_1 \int_{\bar \xi_2}^1 d\xi_2 h(\xi_1,\xi_2)
~\delta\left[\left(\frac{\xi_1}{x_1}-1\right) \left(\frac{\xi_2}{x_2} - 1\right) - \frac{Q_T^2}{M_T^2}\right]\,,
\label{goodlim}
\ea
where $h(\xi_1,\xi_2)$ contains the hard-scattering matrix element and PDFs, and $M_T=\sqrt{Q^2 + Q_T^2}$.
The energy constraint from the $\delta$-function imposes 
the following boundaries on the partonic momentum fractions: 
$\xi_1 =x_1 +(Q_T^2/s)/(\xi_2 -x_2)$; $\bar \xi_1 =[x_1 +(Q_T^2/s)/(1 -x_2)]\leq \xi_1 \leq 1$;
$\bar \xi_2 \equiv [x_2 +(Q_T^2/s)/(1 -x_1)]\leq \xi_2 \leq 1$, with
$x_{1,2}=\frac{M_T}{\sqrt{s}}e^{\pm y}$. 

These boundaries are absent in the $W$ 
and asymptotic contributions, which depend on convolutions of
Wilson coefficient functions and PDFs, 
\ba
&&\left[C_{j,a}\otimes f_{a/h_i}\right](\chi_i,\mu_F)=\int_{\chi_i}^1\frac{d\xi_i}{\xi_i} 
C_{j,a}\left(\frac{\chi_i}{\xi_i},\mu_F
b,C_1,C_2,C_3\right)f_{a/h_i}(\xi_i,
\mu_F)\,
\ea
for $i=1$ or $2$. The variables $\chi_i$ satisfy $\chi_{1,2}
\rightarrow x_{1,2}^{(0)}\equiv (Q/\sqrt{s}) e^{\pm y}$ and cannot
exceed $\bar \xi_{1,2}$. Thus, for non-negligible
$Q_T^2/Q^2$, the $W$ and asymptotic term may include contributions
from the unphysical momentum fractions $\xi_i\leq \bar \xi_i$, 
and ideally one should include kinematically important $Q_T^2/Q^2$
contributions into $\chi_{1,2}$ to bring them as close to $\bar
\xi_{1,2}$ as possible.

As the procedure for including the $Q_T^2/Q^2$ corrections in the $W$
term is not unique, we explored several of them. 
We find that either $\chi_{1,2} = x^{(0)}_{1,2}=(Q/\sqrt{s})e^{\pm y}$
or $\chi_{1,2} = x_{1,2} = (M_T/\sqrt{s})e^{\pm y}$
results in the comparable agreement with the $\phi^*_\eta $ data from
D\O\ and ATLAS 7 TeV. 
These prescriptions are designated as the ``kinematical corrections
of type 0'' and ``type 1'', or $kc_0$ and $kc_1$, in our numerical outputs. 

In contrast, some alternative choices produce worse
agreement with the examined data, such as
$\chi_{1,2} = \bar \xi_{1,2} = ((M_T+Q_T)/\sqrt{s})e^{\pm
  y}$ designated as $kc_2$. Furthermore, the $kc_1$ prescription 
improves matching compared to $kc_0$ at $\sqrt{s}=14 $ TeV,
corresponding to scattering at smaller $x$. We use the $kc_1$ matching
as the default prescription in the subsequent comparisons. 

Dependence on the matching prescription at intermediate $Q_{T}$ (intermediate
$\phi_{\eta}^{*}$) reflects residual sensitivity to higher-order
contributions and is reduced~\cite{Balazs:1997xd} once large-$Q_{T}$
contributions of ${\cal O}(\alpha_{s}^{2})$ are included, compared
to ${\cal O}(\alpha_{s})$. The \textsc{ResBos} implementation follows a general
argument for matching of the resummed contribution onto the fixed-order
result that applies in other areas, such as the treatment of PDFs
for heavy quarks in DIS in a general-mass variable number scheme~\cite{Tung:2001mv,Guzzi:2011ew}.
Matching is stabilized by constructing resummed coefficient functions
that comply with the energy-momentum conservation in the exact fixed-order
contribution.

\subsection{Photon radiative contributions \label{sec:Photon}}

Our resummed calculations include both $Z$-mediated and photon-mediated
contributions to production of Drell-Yan pairs, as well as their
interference. Electroweak radiative contributions 
have been extensively studied in $Z$ boson~\cite{Baur:2001ze,Zykunov:2005tc,CarloniCalame:2007cd,Arbuzov:2007db}
and $W$ boson production~\cite{Dittmaier:2001ay,Baur:2004ig,Cao:2004yy,Zykunov:2006yb,Arbuzov:2005dd,CarloniCalame:2006zq}. 
The dominant NLO electroweak contribution 
is associated with final-state radiation of photons.
To compare the D\O~ data to the \textsc{ResBos} prediction 
without the NLO electroweak correction, we correct
the fitted data to the Born level for final-state leptons 
by subtracting the NLO EM correction obtained bin-by-bin by 
the \textsc{Photos} code~\cite{Barberio:1993qi}. This correction is
essential for the agreement of \textsc{ResBos} theory and
data. However, since the photon-mediated and final-state photon
radiation contributions are relatively small, in the first
approximation we can treat them as a linear perturbation and 
evaluate for a fixed combination of the nonperturbative 
and scale parameters taken either from the BLNY 
or our best-fit parametrizations.

\subsection{Numerical accuracy \label{sec:Num_acc}}

Given the complexity of the resummation calculation, we expect several
sources of random numerical errors that may compete 
with the accuracy of the most precise $\phi^*_\eta$ data points, 
which are of order 0.5\% of the respective central cross sections. 
The numerical errors may arise from the parametrizations of PDFs, 
integration, and interpolation at various stages of the analysis. 
They can be treated as independent and uncorrelated and primarily 
result in higher-than-normal values of the figure-of-merit 
function $\chi^2$ when not explicitly included in the estimates. In
comparison, the variations due to $a_Z$ or $C_{1,2,3}$ parameters 
are of order a few percent and correlated across the $\phi_\eta^*$ spectrum.

\subsection{A comparison with an alternative formalism \label{sec:CSS_vs_BCFG} }

In the last part of this section, it is instructive to summarize the distinctions between
the NNLL/NNLO resummed $Q_T$ distributions obtained in the CSS formalism 
and the alternative approach 
of Refs.~\cite{Bozzi:2005wk,Bozzi:2008bb,Catani:2012qa}. 
Both methods predict $Q_T$ distributions in a wide range of processes, 
including production of lepton pairs and Higgs boson. 
While sharing the same physics principles, they
organize the small-$Q_T$ form factor $W$ 
in distinct ways and differ in the form of their higher-order
corrections and quantitative dependence on QCD scales.

We will outline the key differences by referring 
to the work by Bozzi, Catani, de Florian, and Grazzini (BCFG) in 
Ref.~\cite{Bozzi:2005wk}. There, the BCFG representation was derived 
step-by-step and compared to the CSS method in the kin process of 
$gg\rightarrow\mbox{Higgs}$ production. The general observations of 
that paper also apply to the Drell-Yan process.

In both formalisms, the resummed $Q_{T}$ distribution
for $h_{1}h_{2}\rightarrow VX$, where $V=\gamma^{*},Z$, and upon integration
over the decay angles of the lepton pair, is constructed from the small-$Q_T$ resummed, large-$Q_T$ fixed-order, and asymptotic (overlap) contributions, denoted as $W$, FO, and $(W)_{\mbox{FO}}$:  
\begin{equation}
\frac{d\sigma\left(h_{1}h_{2}\rightarrow VX\right)}{dQ^{2}\ dy_{Z}\ dQ_{T}^{2}}=W+\mbox{FO}-\left(W\right)_{\mbox{FO}}=W+Y.
\end{equation}
In accord with the preceding discussion, the $W$ term of the
CSS formalism takes form, in a simplified
notation, of
\begin{equation}
W(Q,Q_{T},y_{Z})=\int\frac{bdb}{4\pi}\, J_{0}(Q_{T}b)\widetilde{W}_{CSS}(b,Q,y_{Z}).
\end{equation}
The integrand consists of the zeroth order Bessel function, 
$J_0(Q_Tb)$, and the form factor $\widetilde{W}_{CSS}$ that is derived 
in the context of TMD factorization. 
At $b\ll1\mbox{ GeV}^{-1}$, the form factor is expressed as
\begin{eqnarray}
& & \widetilde{W}_{CSS}(b,Q,y_{Z})=\sum_{j,a,b}\int_{0}^{1}d\xi_{1}\int_{0}^{1}d\xi_{2}\ f_{a/h_{1}}(\xi_{1},\mu_{F})\, f_{b/h_{2}}(\xi_{2},\mu_{F})\nonumber \\
& \times & \left|H_{j}(Q,\mu_{Q}/Q;\alpha_{S}(\mu_{Q}))\right|^{2}\,\exp\left[-\int_{\mu_{b}}^{\mu_{Q}}\frac{d\mu^{2}}{\mu^{2}}A(\alpha_{S}(\mu),\,\mu_{b}b)\,\ln\left(\frac{\mu_{Q}^{2}}{\mu^{2}}\right)+B(\alpha_{S}(\mu),\,\mu_{b}b,\,\mu_{Q}/Q)\right]\nonumber \\
& \times & 
{\cal C}_{ja}\left(\frac{\chi_{1}}{\xi_{1}},\frac{\mu_{Q}}{\mu_{b}},\mu_{F}b\right){\cal C}_{\bar{j}b}\left(\frac{\chi_{2}}{\xi_2},\frac{\mu_{Q}}{\mu_{b}},\mu_{F}b\right).
\label{WCSS}
\end{eqnarray}
$\widetilde{W}_{CSS}$ depends on three QCD scales: $\mu_{b}=C_{1}/b$,
$\mu_{Q}=C_{2}Q$, $\mu_{F}=C_{3}/b$, where the arbitrary scale constants
$C_{1},$ $C_{2}$, and $C_{3}$ are of order unity.
Their exact values are chosen so as to optimize the convergence of the perturbative
series. The combination $C_1=C_3=b_0$, $C_2=1$ is the natural choice.

In Ref.~\cite{Bozzi:2005wk,Bozzi:2008bb}, the resummed form factor 
$\widetilde{W}_{BCFG}$ in the second approach is written at $b\ll1\mbox{ GeV}^{-1}$ as 
\begin{align}
\widetilde{W}_{BCFG}(b,Q,y_{Z}) = &\nonumber \\
\sum_{j,a,b}\int_{0}^{1}d\xi_{1}\int_{0}^{1}d\xi_{2}\, & f_{a/h_{1}}(\xi_{1},\overline{\mu}_{F})\, f_{b/h_{2}}\xi_{2},\overline{\mu}_{F})\, W_{j,ab}(b,Q,\xi_{1}\xi_{2}s;\overline{\alpha}_s,\overline{\mu}_{R},\overline{\mu}_{F}).\label{WBCFG}
\end{align}
$W_{j,ab}$ is reconstructed from its $N$-th Mellin moments $W_{ab,N}$ that are expanded in powers 
of $\alpha_s(\overline{\mu}_R)\equiv \overline{\alpha}_s$. 
$W_{ab,N}$ consists of the function ${\cal H}_{j,ab,N}$ that depends only on
scales of order $Q$, and the exponent $e^{{\cal G_{N}}}$ that depends
on $\widetilde{L}\equiv \ln(Q^{2}b^{2}/b_{0}^{2}+1)$ and ratios of various scales:
\begin{align}
& W_{ab,N}\left(b,Q;\overline{\alpha}_s,\overline{\mu}_{R},\overline{\mu}_{F}\right)= \nonumber\\
& \quad \quad {\cal H}_{j,ab,N}\left(Q,\overline{\alpha}_s,Q/\overline{\mu}_{R},Q/\overline{\mu}_{F},Q/\overline{\mu}_{Q}\right)
\cdot
\exp\left\{{\cal G}_{N}\left(\overline{\alpha}_s,\widetilde{L};Q/\overline{\mu}_{R},Q/\overline{\mu}_{Q}\right)\right\}.
\label{WBCFGN}
\end{align}
On the right-hand side, the representation includes three auxiliary QCD 
scales, each taken to be of order of the boson's virtuality $Q$: the resummation
scale $\overline{\mu}_{Q}$, the renormalization scale $\overline{\mu}_{R}$,
and the PDF factorization scale $\overline{\mu}_{F}$. 
The dependence on $b$ enters only through the 
logarithmic term $\widetilde{L}$ inside $e^{{\cal G_{N}}}$.

The representation $\widetilde{W}_{BCFG}(b,Q,y_{Z})$
in Eqs.~(\ref{WBCFG}, \ref{WBCFGN}) can be obtained
from $\widetilde{W}_{CSS}(b,Q,y_{Z})$ in Eq.~(\ref{WCSS}) 
by a series of steps that are documented in~\cite{Bozzi:2005wk}.

First, the QCD scales are selected differently in the two approaches. 
In $\widetilde{W}_{CSS}(b,Q,y_{Z})$ several terms depend on the variable scales
$\mu_{b}=C_1/b$ and $\mu_{F}=C_3/b$. The QCD scale 
$C_{2}Q$ plays the role that is similar 
to the resummation scale $\overline{\mu}_{Q}$. Inside the Sudakov
integral, the scale $\mu$ in $\alpha_s(\mu)$ is integrated over. 

In $\widetilde{W}_{BCFG}(b,Q,y_{Z})$, 
the scales $\overline{\mu}_b$ and $\overline{\mu}_F$ 
are fixed at $b_0/b$. The 
QCD coupling strength $\alpha_s(\mu)$ is converted into the series 
of $\overline{\alpha}_s$ (at the scale $\overline{\mu}_R \sim Q$) 
using the renormalization group equations. The collinear PDFs 
$f_{a/h}(\xi, \mu)$ in Eqs.~(\ref{WCSS}) and (\ref{WBCFG}) are evaluated 
at $\mu_{F}\sim1/b$ in $\widetilde{W}_{CSS}(b,Q,y_{Z})$, and
$\overline{\mu}_{F}\sim Q$ in $\widetilde{W}_{BCFG}(b,Q,y_{Z})$.
To preserve the factorization scale invariance, the Mellin moment $W_{ab,N}$
of the BCFG form factor explicitly includes
an operator matrix ${\bold U}_N(b_0/b,\overline{\mu}_F)$ for DGLAP 
evolution of $f_{a/h}(\xi, \mu)$ 
between the scales $\overline{\mu}_F$ and $b_0/b$, while the CSS form factor does not.\footnote{More specifically, $W_{ab,N}$ in Appendix A of~\cite{Bozzi:2005wk} includes the evolution operator that is factorized as 
${\bold U}_N(b_0/b,\overline{\mu}_F)={\bold U}_N(b_0/b,\overline{\mu}_Q){\bold U}_N(\overline{\mu}_Q,\overline{\mu}_F)$. ${\bold U}_N(b_0/b,\overline{\mu}_Q)$ is exponentiated inside ${\cal G}_{N}$. ${\bold U}_N(\overline{\mu}_Q,\overline{\mu}_F)$ is retained in ${\cal H}_{j,ab,N}$.}

After the conversion $\alpha_s(\mu)\rightarrow
\overline{\alpha}_s$ in $W_{ab,N}$, the
contributions at scales of order $Q$ are included into ${\cal
  H}_{j,ab,N}$ as in Eq.~(\ref{WBCFGN}). The Sudakov integral
$S(b,Q;C_{1},C_{2})=S(\mu_{Q}/\mu_{b})$ and the anomalous dimensions
of various components are assimilated into ${\cal G}_{N}$. 
Within ${\cal G}_{N}$, all evolution operators are expanded as a series in 
$\overline{\alpha}_s$ and $L=\ln(Q^{2}b^{2}/b_{0}^{2})$: 
\begin{equation}
{\cal G}_{N}=Lg^{(1)}(\overline{\alpha}_s L)+
g^{(2)}\left(\overline{\alpha}_s L;
  \frac{Q}{\overline{\mu}_R},
  \frac{Q}{\overline{\mu}_Q}\right) 
+ \frac{\overline{\alpha}_s}{\pi}\,
g^{(3)}\left(\overline{\alpha}_s L;
  \frac{Q}{\overline{\mu}_R},
  \frac{Q}{\overline{\mu}_Q}\right)+... 
\end{equation}

Finally, a prescription 
for matching of the $W+Y$ and FO terms at large $Q_{T}$ 
is introduced in $\widetilde{W}_{BCFG}$ by replacing all
generic logarithms $L=\ln(Q^{2}b^{2}/b_{0}^{2})$ in $e^{{\cal G}_{N}}$
by $\widetilde{L}=\ln(Q^{2}b^{2}/b_{0}^{2}+1)$. The replacement
forces the exponential to satisfy 
$e^{{\cal G_{N}}}\rightarrow 1$ when $b_{0}^{2}/b^{2}\ll Q^{2}$,
i.e., in the region of the small transverse positions $b$ that dominate
the Fourier-Bessel integral when $Q_{T}$ is large. The resulting outcome
is that the $W$ and asymptotic terms cancel well at large
$Q_{T},$ and that, upon the integration over $Q_{T}$, the inclusive
$W+Y$ cross section turns out to be exactly equal to the fixed-order
cross section. In the BCFG cross section, 
matching therefore arises as a result of a mathematical
replacement $L\rightarrow\widetilde{L}$ in the resummed exponential,
and not because 
of the physical constraint due to energy-momentum conservation
imposed in our approach. The $L\rightarrow\widetilde{L}$ matching
works by suppressing the ${\cal G_{N}}$ exponent at $b^{2}\ll1/Q^{2}$
via a deft, even though not unique, redefinition of $L$ (see also Ref.~\cite{Catani:1992ua}). 

In the CSS approach adopted in \textsc{ResBos}, 
the scale constants $C_{1}$ and $C_{3}$ need 
not to equal $b_0$ exactly. The Sudakov integral $S(\mu_{Q}/\mu_{b})$ 
is evaluated numerically and not as a logarithmic expansion 
in powers of $L$ as it is done in $\widetilde{W}_{BCFG}(b,Q,y_{Z})$. The \textsc{ResBos}
code does not operate with the independent QCD scales $\overline{\mu}_R\sim Q$ 
and $\overline{\mu}_F \sim Q$ of the BCFG formalism. 

\textsc{ResBos} finds $\alpha_s(\mu)$ by numerically solving the
renormalization group equation and always evolves 
the PDFs $f_{a/h}(\xi,\mu_{F})$ 
forward from the initial scale $Q_{0}\approx1$ GeV of the
input PDF ensemble to a higher scale $\mu_{F}\geq Q_{0}.$ This is to
be contrasted with $\widetilde{W}_{BCFG}(b,Q,y_{Z})$, which implements the
logarithmic expansion for $\alpha_s(\mu)$
and the DGLAP matrix operator ${\bold
  U}_N(b_0/b,\overline{\mu}_Q)$  that evolves the
PDFs $f_{a/h}(\xi,\overline{\mu}_{F})$ backward from 
$\overline{\mu}_{Q}\approx Q$ down to a lower scale
$b_0/b$ in the most relevant $b$ region.
The backward evolution of this kind has a tendency
to be unstable and cause the PDFs to deviate at 
low momentum scales. Hence the numerical evolution of $\alpha_s(\mu)$
and forward DGLAP evolution adopted in $\widetilde{W}_{CSS}(b,Q,y_{Z})$
is more trustworthy in precision studies. 

The nonperturbative contribution arises in 
$\widetilde{W}_{CSS}(b,Q,y_{Z})$ as a natural feature
of QCD factorization in terms of TMD PDFs.
Dependence on matching is present,
implicitly or explicitly, in either formalism.
When looking for evidence of nonperturbative effects in $\phi_{\eta}^{*}$
distributions, it is desirable to investigate several prescriptions
for matching of low-$Q_{T}$ and high-$Q_{T}$ terms. We have done
it by varying the form of the rescaling variables that control the
cross sections in the matching region in \textsc{ResBos}.

\section{Numerical results \label{sec:numerical-results}}

\subsection{General features}
In this section we determine $a_Z$ from the distribution
$(1/\sigma)~d\sigma/d\phi^*_\eta$ measured by
D\O\ ~\cite{Abazov:2010mk}
that is normalized to the total cross section
$\sigma$ in the measured $Q$ and $y$ range.
These data are given in three bins of $Z$ boson rapidity $y_Z$.
In the first two, $|y_Z|\leq 1$ and $1\leq |y_Z|\leq 2$,
the $(1/\sigma)~d\sigma/d\phi^*_\eta$ distribution 
is measured separately for electrons and muons at $N_{pt}=29$ points
of $\phi^*_\eta$. In the third bin,  $|y_Z|\geq 2$, 
only electrons are measured
at 25 points of $\phi^*_\eta$. The first two $y_Z$ bins provide
substantial new constraints. The third bin has larger statistical
errors and reduced discriminating power. 

All predictions are obtained by using CT10 NNLO PDFs~\cite{Gao:2013xoa}. Predictions based on 
MSTW'08 NNLO PDF sets~\cite{Martin:2009iq} were also
computed and did not show significant difference with  CT10 NNLO predictions.

\begin{figure}[tb]
\begin{center}
\includegraphics[width=13cm, angle=0]{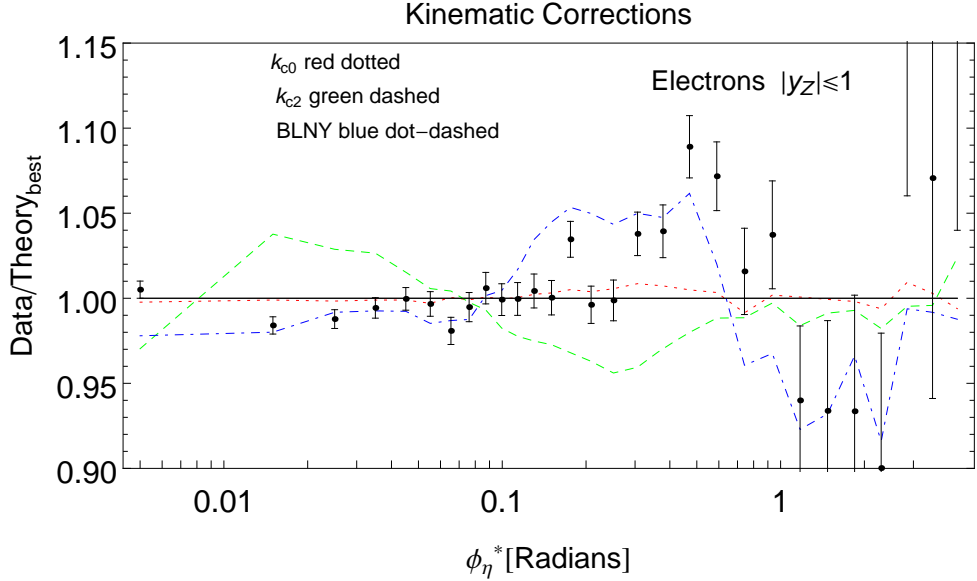}
\caption{The ratios to the central theoretical prediction of 
the D\O\ electron data at $|y_Z|\leq 1$ and alternative theoretical
predictions. The central prediction is computed assuming
$C_1=C_3=2b_0$, $C_2=1/2$, $a_Z=1.1\mbox{ GeV}^2$, and kinematical
correction 1. Theory predictions based on alternative kinematical
corrections (0 and 2) and BLNY nonperturbative parametrization are
also shown.  
\label{fig:kc012BLNY}}
\end{center}
\end{figure}

\begin{figure}[p]
\begin{center}
\includegraphics[width=10cm, angle=0]{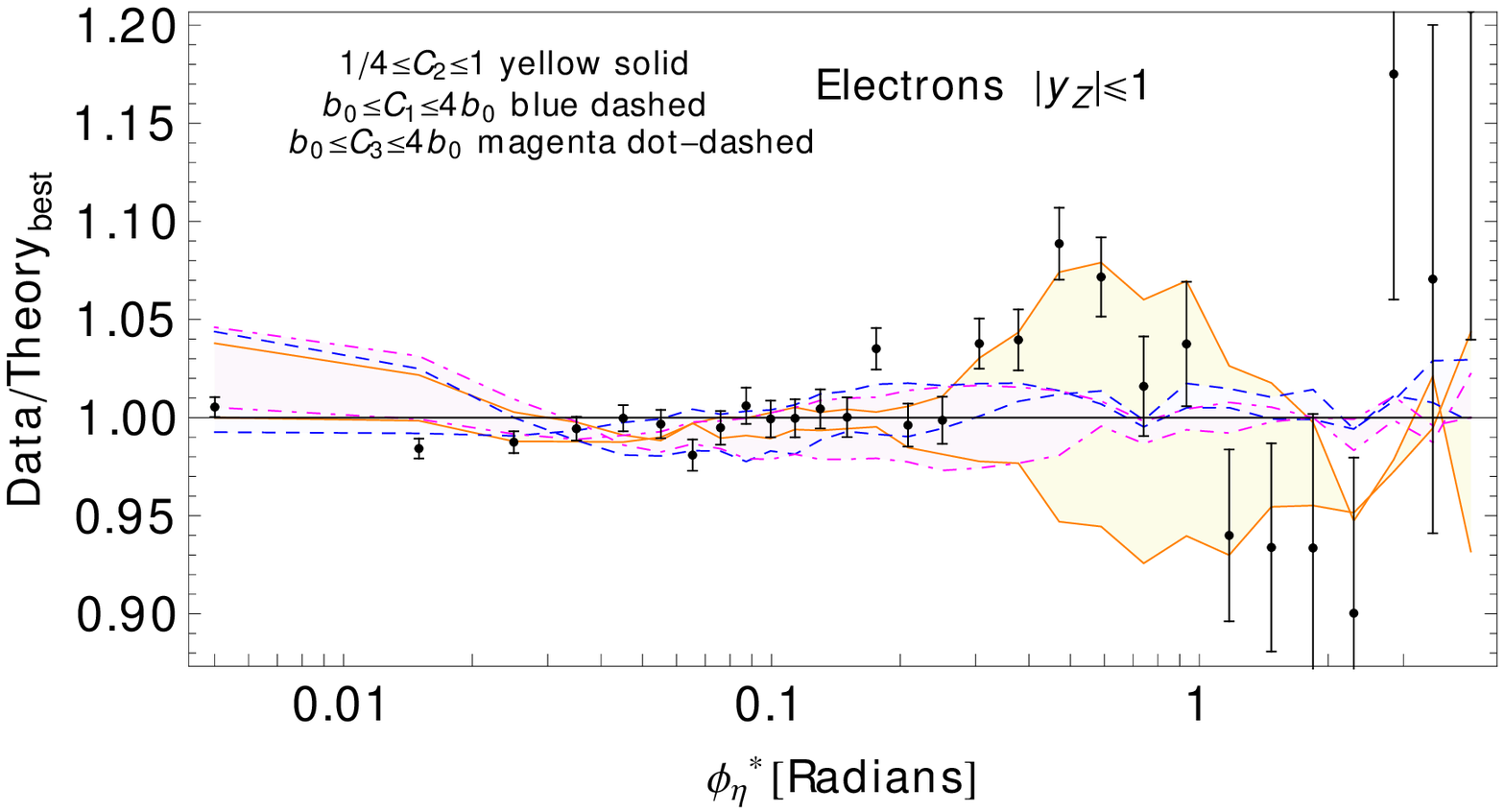}
\includegraphics[width=10cm, angle=0]{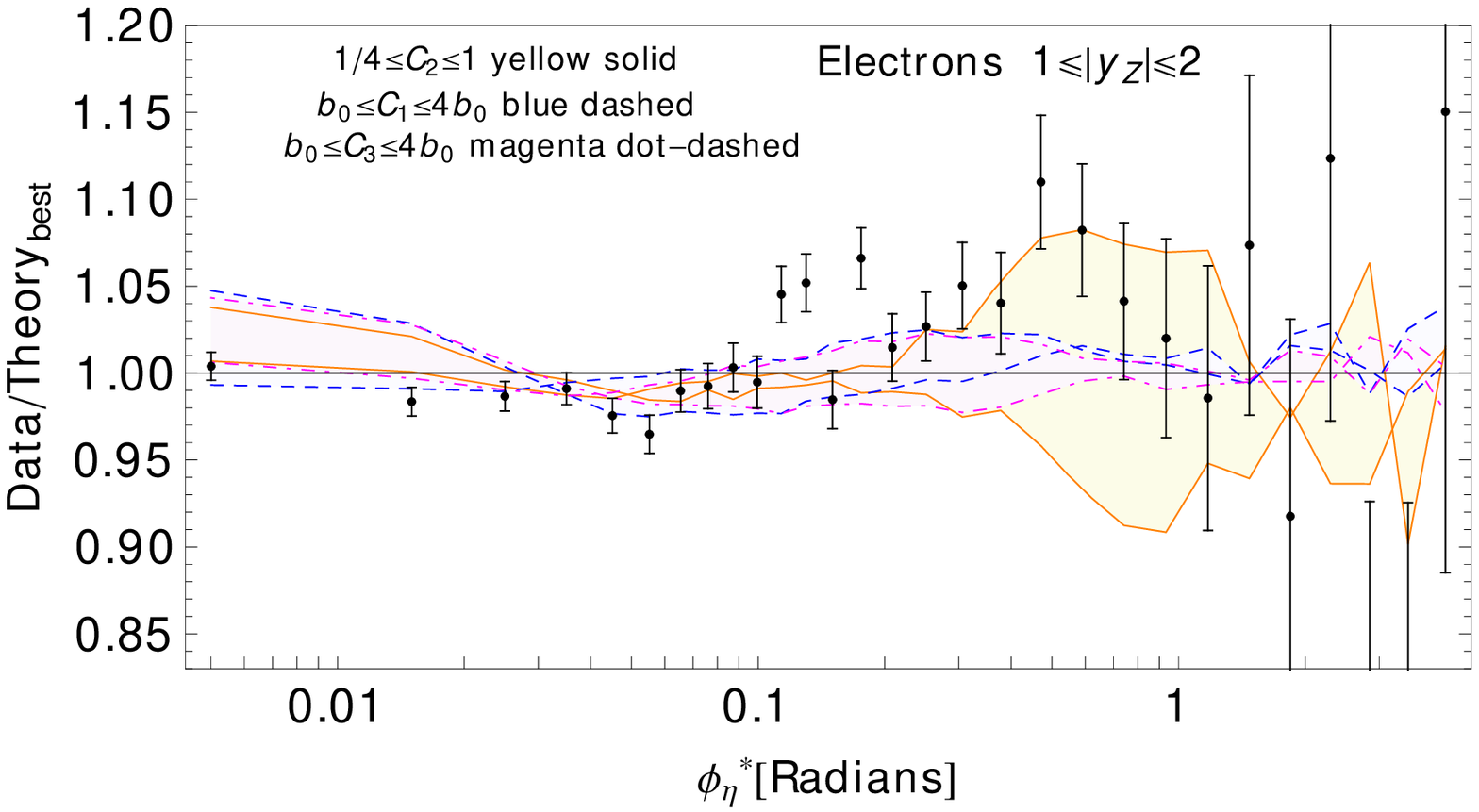}
\includegraphics[width=10cm, angle=0]{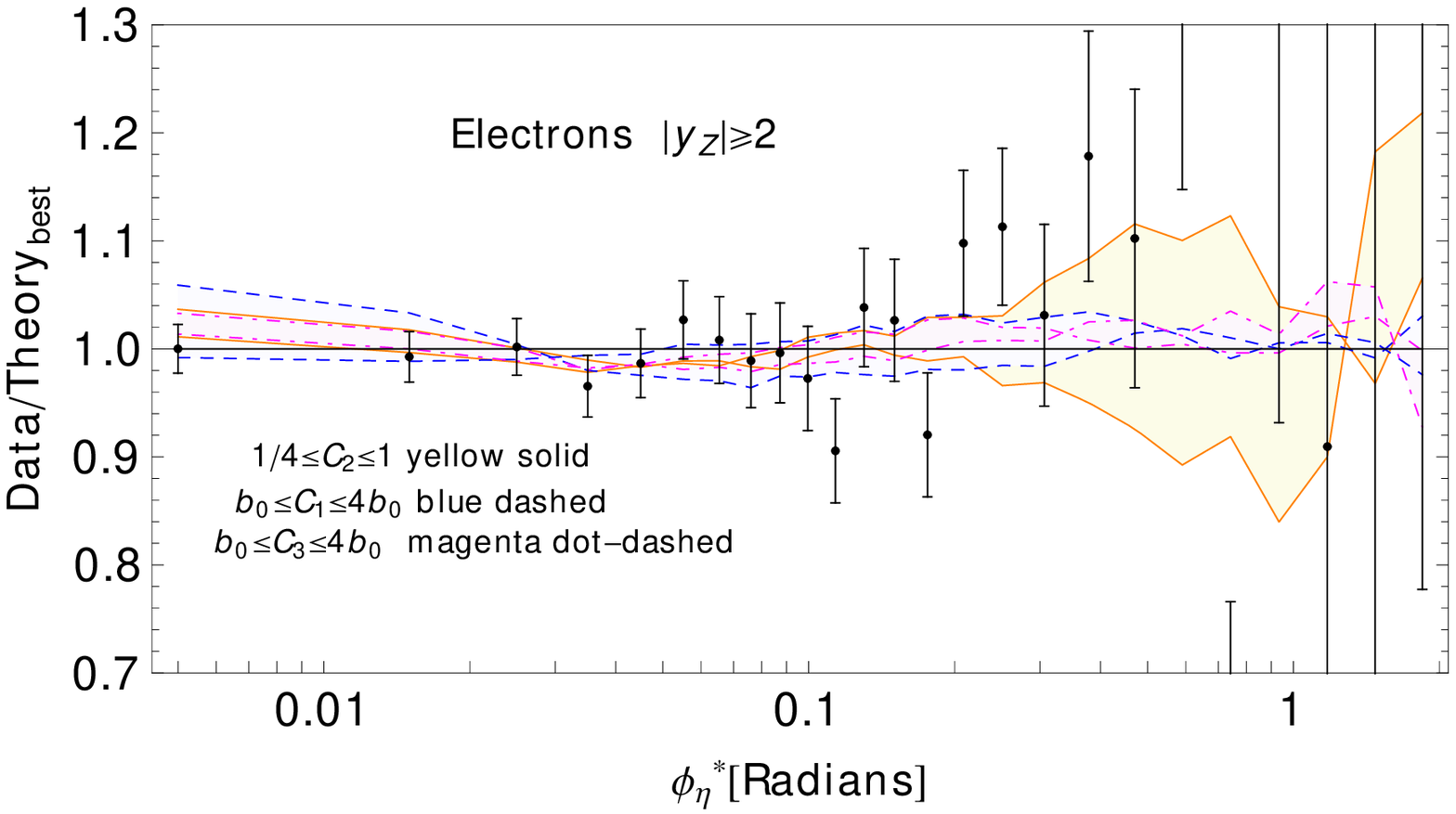}
\caption{Electrons: scale variation due to $C_{1,2,3}$ at small $\phi^*_{\eta}$.
\label{scaleC2E}}
\end{center}
\end{figure}
\begin{figure}[p]
\begin{center}
\includegraphics[width=10cm, angle=0]{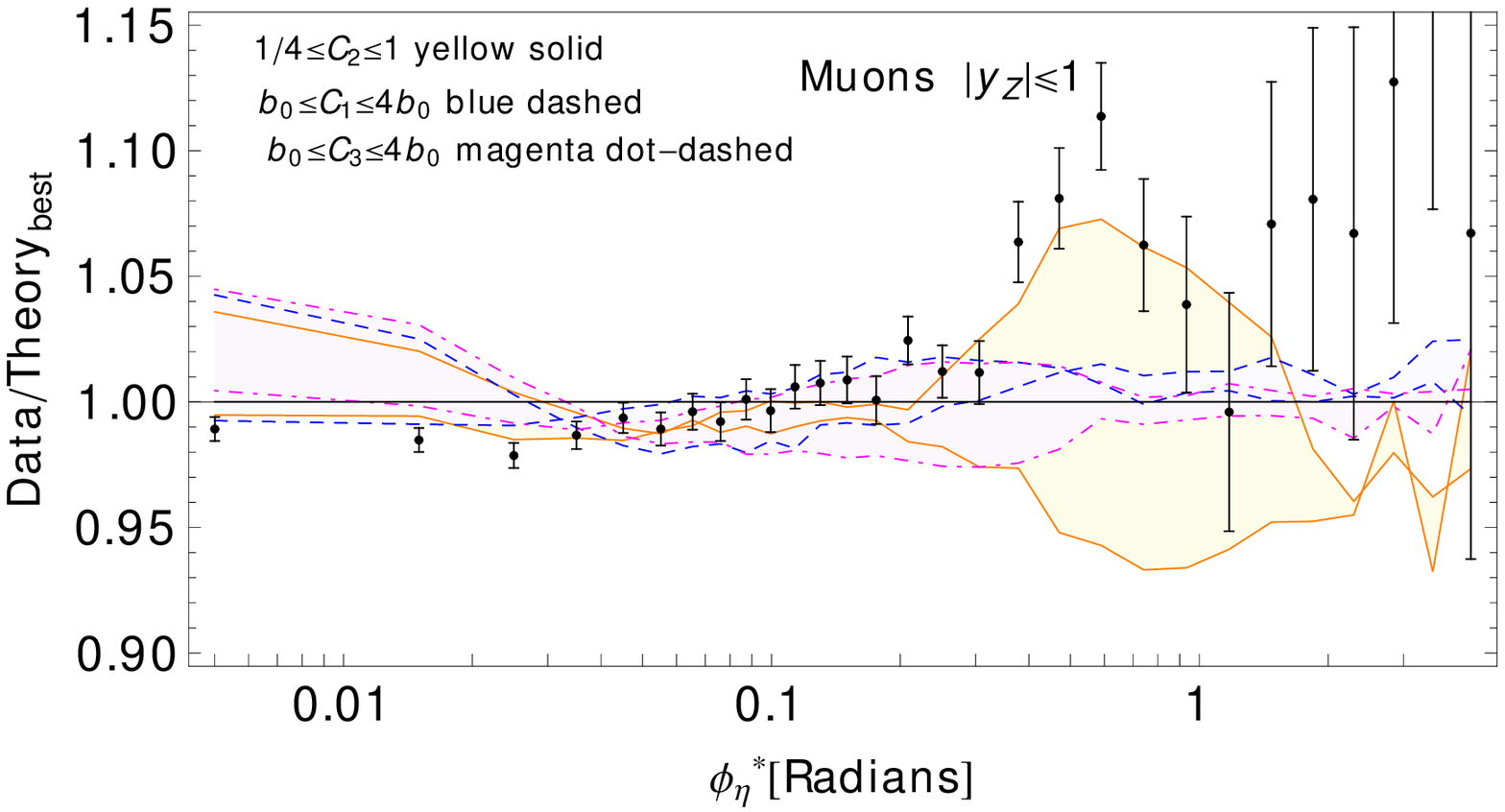}
\includegraphics[width=10cm, angle=0]{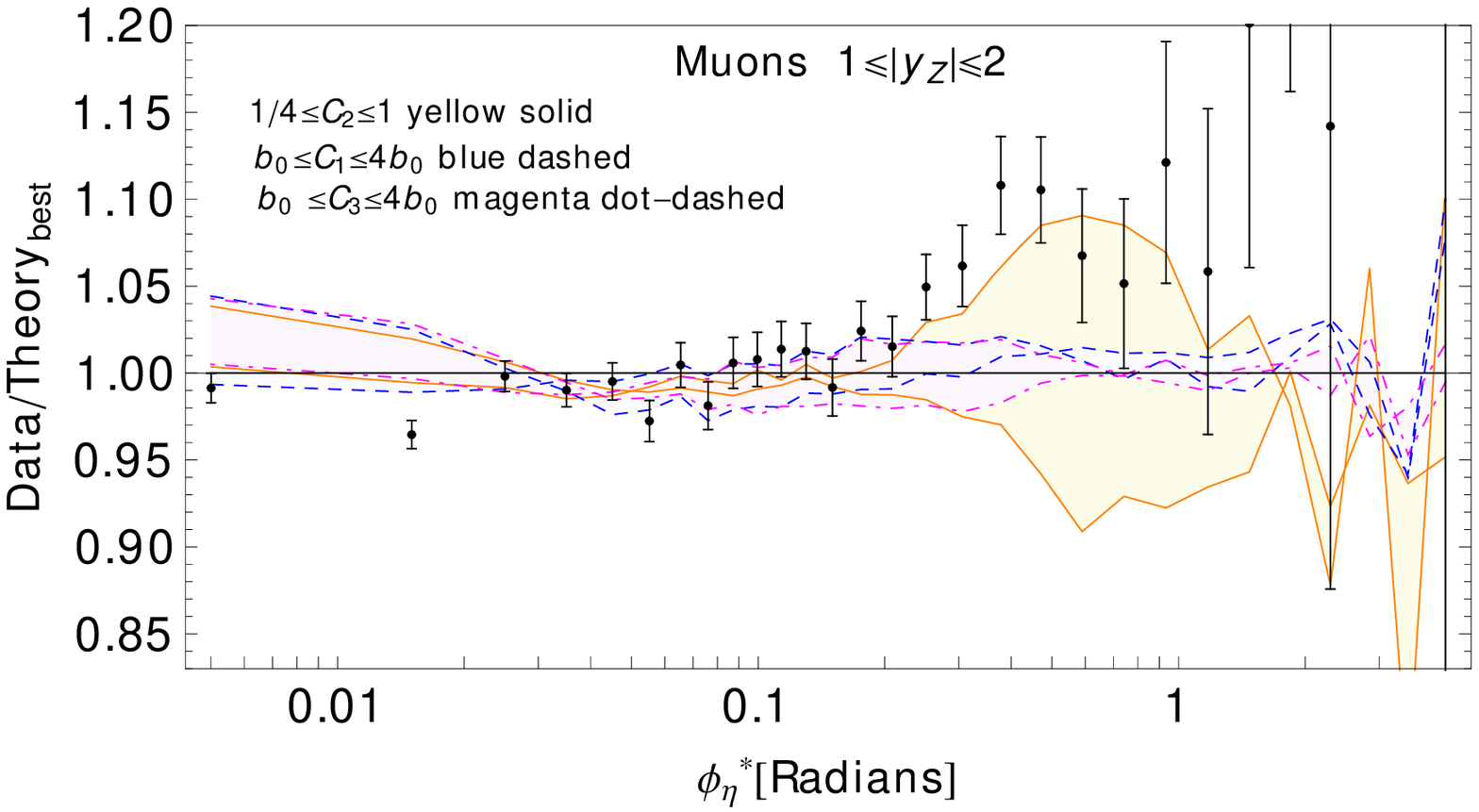}
\caption{Same as above but for the muons.\label{scaleC2M}}
\end{center}
\end{figure}

From the previous section, the resummed cross sections
depend on the perturbative scales, power-suppressed contributions, and
choice of subleading kinematic terms. It is possible to identify an
optimal combination of these factors that results in a good
description of the D\O\ data across the full $\phi^*_\eta$
range. In particular, the large-$Q_T$/large-$\phi_\eta^*$ data
generally prefer the factorization scale of order $Q/2$ or even less 
in the fixed-order piece. At small-$Q_T$(small-$\phi_\eta^*$), 
the scale parameter $C_3$ in the range $1.5b_0-2b_0$ is slightly more
preferable. To illustrate properties of the $\phi_\eta^*$
distributions, we compute the resummed
cross sections using a combination $C_1=C_3=2\ b_0$, $C_2=Q/2$, 
and $a_Z=1.1\mbox{ GeV}^2$ that is close to the best-fit solution. 
The difference between the best-fit solution and the prediction based
on these round-off values will be discussed in Sec.~\ref{sec:chi2analysis}. 

A comparison of the prediction with these choices to the D\O\ data for
$|y_Z|\leq 1$ and a few other predictions is presented in
Fig.~\ref{fig:kc012BLNY}. The new parametrization provides better
description of the data at $0.1\leq \phi^*_\eta \leq 1$ than the
superimposed prediction utilizing the BLNY parametrization~\cite{Landry:2002ix} of $\widetilde{W}^{NP}$. Consequently, it results in
a better $\chi^2$ than the \textsc{ResBos} prediction 
used in the D\O\ analysis~\cite{Abazov:2010mk}, which used 
the CTEQ6.6 NLO PDFs, BLNY $\widetilde{W}_{NP}$, and canonical choice of
$C_{1,2,3}$.

We also compare predictions
for three types (0, 1, 2) of the kinematical (matching) correction
discussed in Sec.~\ref{sec:Matching}. For the selected combination of
scale parameters, the type-0 and 1 kinematical corrections provide a
nearly identical prediction. The type-0 and type-1 corrections can
differ by 2-3\% for other scales. Type 2 is generally disfavored, so
that we assume the type-1 correction for the rest of the analysis.  

A prediction with the same theoretical parameters, as well as for
variations in QCD scales in the ranges $1/4\leq C_2\leq 1$ and
$b_0\leq C_{1,3}\leq 4b_0$, 
are compared to the data for electron production in Fig.~\ref{scaleC2E}
and muon production in Fig.~\ref{scaleC2M}. Here we show all 
rapidity bins both for electron and muon samples. 
The ratios of the  D\O\ data to \textsc{ResBos} theory with the optimal
parameters are indicated by  black circles. Yellow solid, blue dashed, 
and magenta dot-dashed bands represent variations in theory due to
$C_2$, $C_{1}$, and $C_{3}$, respectively, all normalized to the best-fit
prediction. Again, the agreement with \textsc{ResBos} 
observed in these figures is
better than in~\cite{Abazov:2010mk}.
Figs. \ref{scaleC2E} and \ref{scaleC2M} demonstrate that the theoretical 
uncertainty at small $\phi^*_\eta$ is dominated by variations 
of $C_1$ and $C_3$. The bands of scale uncertainty 
are reduced significantly for $0.04\leq\phi^*_{\eta}\leq 0.1$ upon the
inclusion of ${\cal O}(\alpha_s^2)$ scale dependence, as has been
discussed in Sec.~\ref{sec:WpertAS}.

\begin{figure}
\begin{center}
\includegraphics[width=12cm, angle=0]{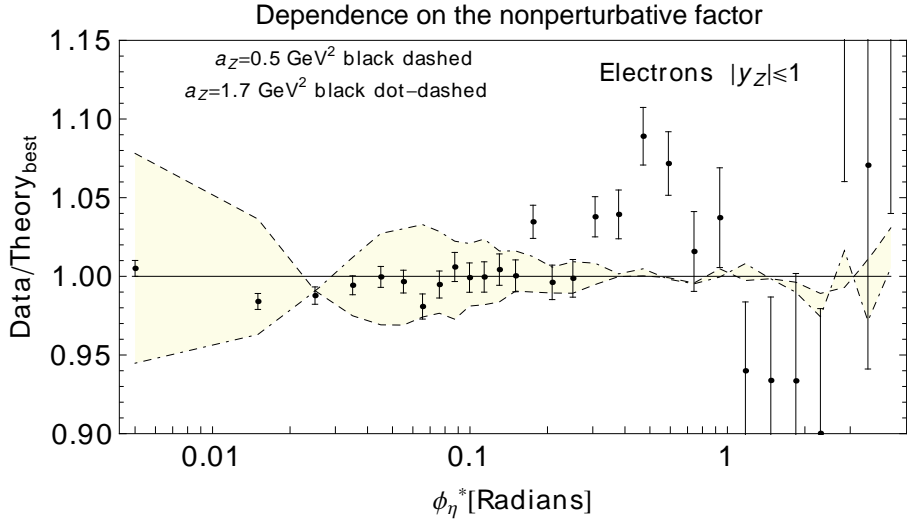}

\caption{Dependence on the nonperturbative parameter $a_Z$ 
for electrons with $|y_Z|\leq 1$.\label{fig:aZkc1}}
\end{center}
\end{figure}

The scale variations can be compared to the dependence on $a_Z$
and kinematic correction in Fig.~\ref{fig:aZkc1}, which result in 
distinctly different patterns of varation in
$d\sigma/d\phi^*_\eta$. In particular, while the perturbative
scale coefficients $C_1, C_2, C_3$ produce a slowly changing
variation across most of the measured $\phi^*_\eta$ range, the
increase in $a_Z$ produces a distinct variation that suppresses
the rate at $\phi^*_\eta \lesssim 0.02$ and increases it at $0.02
\lesssim \phi^*_\eta \lesssim 0.5$, with the rate above $0.5$
essentially unaffected. 
A similar behavior was observed in Fig. 6 of~\cite{Banfi:2011dm}
with a different NP function and a different procedure that 
varied all QCD scales around the central values of order of 
the dilepton's mass.

It is therefore possible to separate the scale dependence from the $a_Z$
dependence if we restrict the attention to $\phi^*_\eta$ 
below and around $\phi^*_\eta=0.1$. To this aim we consider 
only the first 12 bins of $\phi^*_\eta$, starting from the smallest
value, for each value of rapidity. Extending 
the fitted range above $\phi^*_\eta\geq 0.1$ has a minimal effect on $a_Z$.

\subsection{Detailed analysis\label{sec:chi2analysis}}
We pursue two approaches for the examination of the
low-$\phi^*_\eta$ region. In method I, we study the dependence on
$a_Z$ by assuming fixed resummation scales 
corresponding to half-integer scale parameters, 
such as $C_1/b_0=C_3/b_0=1, C_2=1/2$ or
$C_1/b_0=C_3/b_0=2,C_2=1/2$. In this method, 
the goodness-of-fit function $\chi^2$ is minimized with respect to $a_Z$ 
for select combinations of fixed scale parameters.
We find that a $\chi^2$ minimum with respect to
$a_Z$ exists in these cases, but, given 
the outstanding precision of the $\phi^*_\eta$ data,
the best-fit $\chi^2/N_{pt}$ remains relatively high, of order 2-3. This is partly due to the numerical noise
discussed in Sec.\ref{sec:Num_acc}.

The $\chi^2$ function can be further reduced by allowing arbitrary
$C_{1,2,3}$ parameters, in particular, by taking $C_2$ to
be {\it below} 1/2. 
In this context, one has to decide on the acceptable range of
variations in  $C_{1,2,3}$, {\it i.e.} the resummation scales.

As computations for multiple combinations of $a_Z$  and $C_{1,2,3}$ parameters
would be prohibitively CPU-extensive, in method II we first 
consider a fixed scale combination indicated 
by $\{\bar{C}_1, \bar{C}_2, \bar{C}_3\}$ and 
implement a linearized model for small deviations of the scale parameters
from $\bar{C}_{1,2,3}$. The central combination $\bar{C}_{1,2,3}$,
namely ${\bar C}_1 = {\bar C}_3=2 b_0$, ${\bar C}_2=1/2$, 
produces good agreement with the data, although not as good 
as completely free $C_{1,2,3}$. The linearized model is explained 
in Sec.~\ref{methodII}. It provides a fast
estimate of small correlated changes in the $\phi^*_\eta$ shape 
of the kind shown in Figs.~\ref{scaleC2E} and \ref{scaleC2M}. 

The $\chi^2$ function is sampled at discrete $a_Z$ values in the
interval  $a_Z=[0.1:3.5]$ GeV$^{2}$ and reconstructed between 
the sampling nodes by using polynomial interpolation. 
When the scale variations are allowed, the
dependence of $\chi^2$ on $a_Z$ is asymmetric and 
very different from a quadratic one. 
 
To account for the asymmetry of the distributions, we quote the central value 
$\overline{a}_Z$ that minimizes
$\chi^2(a_Z)$ and the 68\% confidence level (C.L.) uncertainty.
The probability density function ${\cal P}(a_Z)$ 
for $a_Z$ in a sample with $N$ points is taken to follow a chi-squared
distribution  with $N$ degrees of freedom, 
\beq
{\cal P}(a_Z)= {\cal
  P}_{\chi}(N,\chi^2(a_Z))=\frac{(\chi^2)^{N/2-1}\exp\left(-\chi^2/2\right)}{\Gamma(N/2)\ 2^{N/2}}.
\eeq
With this, we determine the 68\% C.L. intervals $[a_{Z,min},~a_{Z,max}]$, 
where $a_{Z,min}$ and $a_{Z,max}$ are defined implicitly by 
\ba
0.16=\frac{\int^{a_{Z,min}}_{0} {\cal P}(a_Z)~d a_Z}{\int^{+\infty}_{0} {\cal P}(a_Z)~d a_Z}\,,
&&
0.84=\frac{\int^{a_{Z,max}}_{0} {\cal P}(a_Z)~d a_Z}{\int^{+\infty}_{0} {\cal P}(a_Z)~d a_Z}\,.
\label{aZmin_max}
\ea
For an asymmetric distribution as in method II, the central value
$\overline{a}_Z$ does not coincide with the middle of the 68\%
C.L. interval or the mean $a_Z$ given by the first moment of the
${\cal P}(a_Z)$ distribution.

\subsubsection{Method I: minimization with fixed scale parameters}
\label{methodI}

\begin{figure}[tb]
\begin{center}
\includegraphics[width=0.49\textwidth]{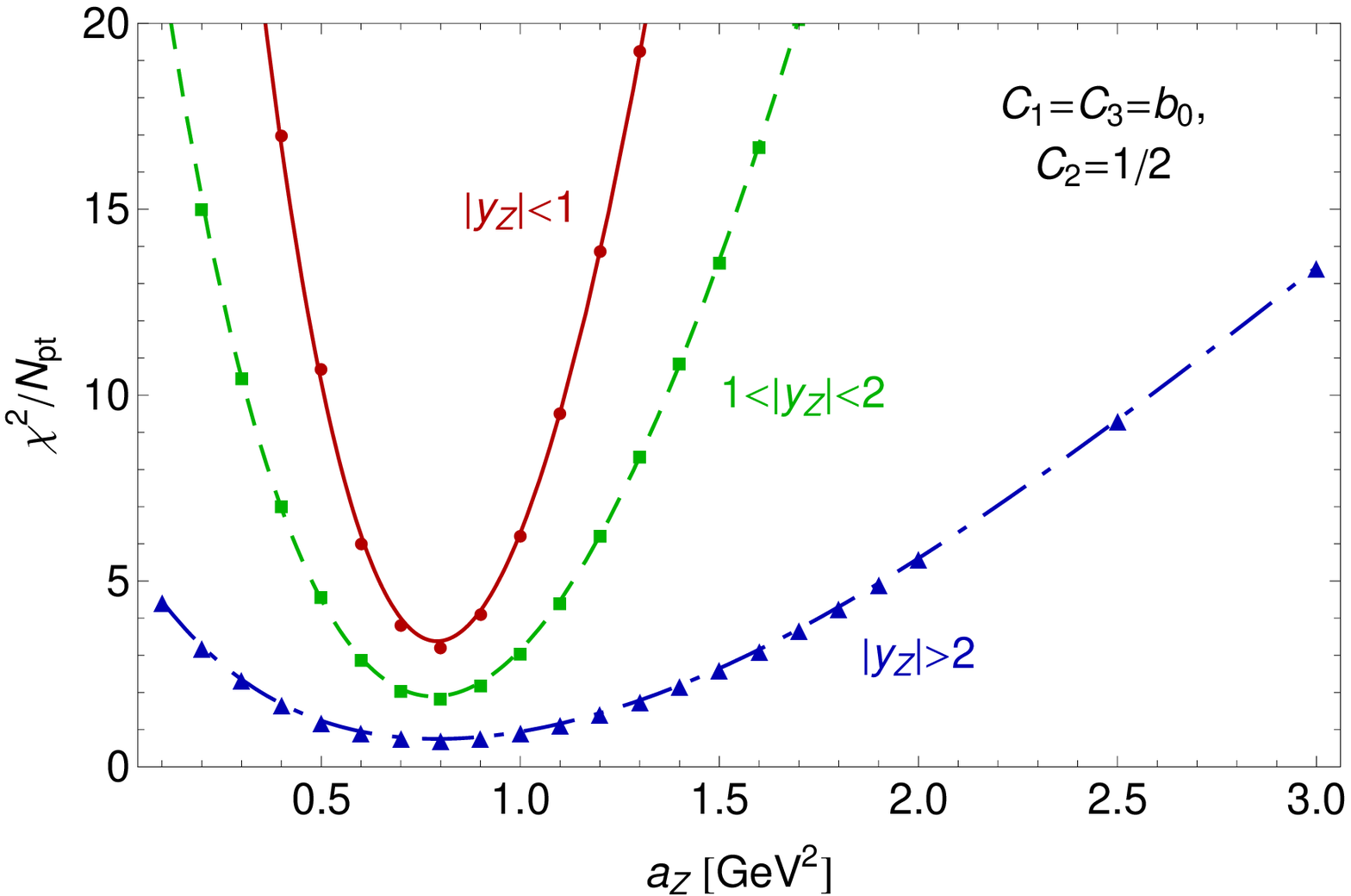}
\includegraphics[width=0.49\textwidth]{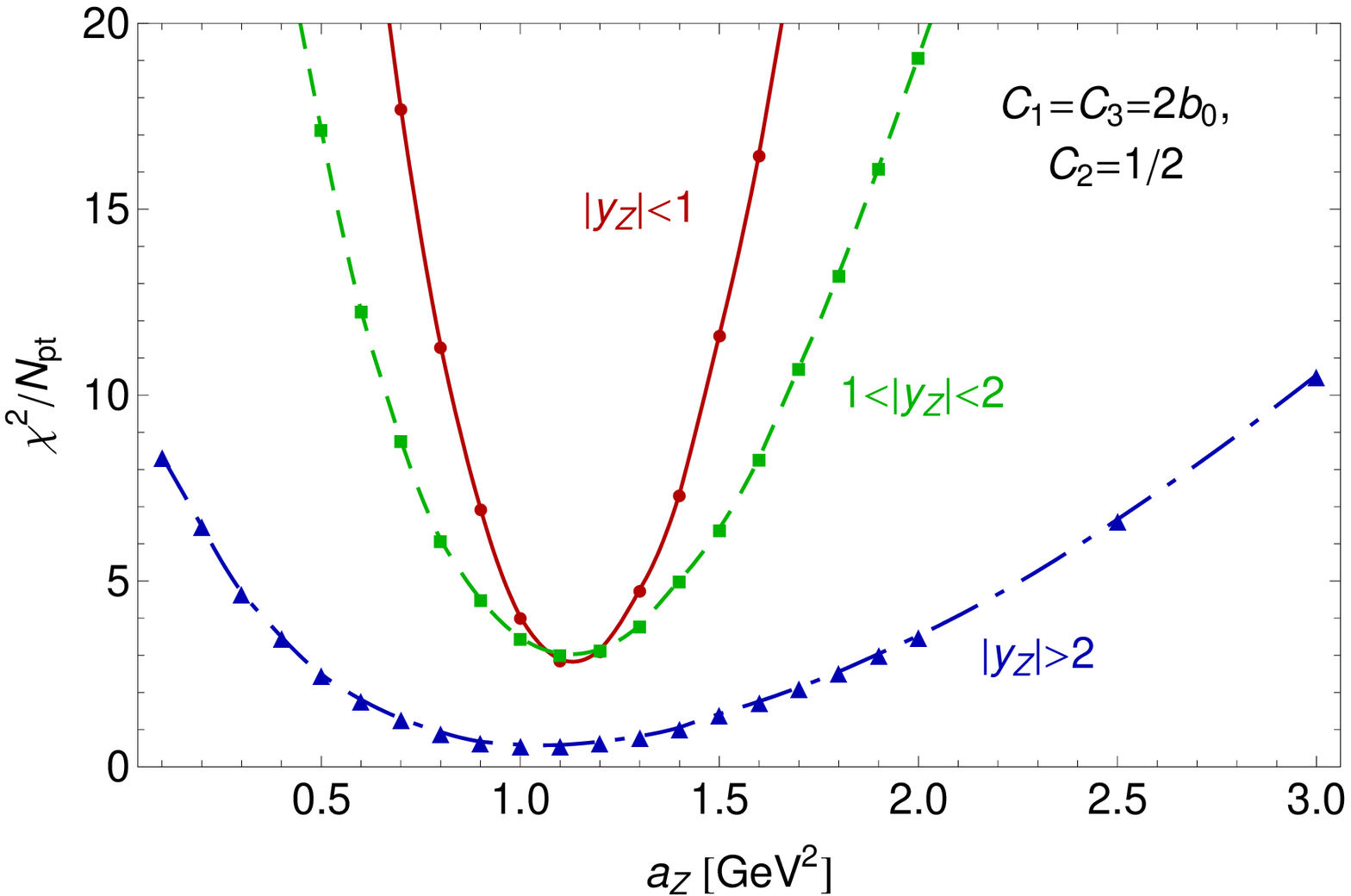}
\caption{$\chi^2/N_{pt}$ as a function of $a_Z$ with fixed $C_{1,2,3}$.
\label{chivsa1}}
\end{center}
\end{figure}

\begin{table}
\centering{}%
\begin{tabular}{|c|c|c|c|}
\hline 
\multicolumn{4}{|c|}{ Fit results for $\phi_{\eta}^{*}\leq0.1$}\tabularnewline
\hline 
\hline 
 & $N_{pt}$  & $\chi_{min}^{2}/N_{pt}$  & $\overline{a}_{Z}\pm\delta a_{Z}\ (\mbox{GeV}^{2})$ \tabularnewline
\hline 
$|y_{Z}|\leq1$, ~~$e+\mu$  & 24 & 3.24  & $0.79_{-0.03}^{+0.2}$ \tabularnewline
 &  & 2.83  & $1.14\pm0.08$ \tabularnewline
\hline 
$1\leq|y_{Z}|\leq2$,~~ $e+\mu$  & 24 & 1.87  & $0.79\pm0.05$ \tabularnewline
 &  & 3.03  & $1.12_{-0.13}^{+0.14}$ \tabularnewline
\hline 
$|y_{Z}|\geq2$, ~~$e$ & 12 & 0.74 & $0.8_{-0.05}^{+0.03}$\tabularnewline
 &  & 0.58 & $1.04_{-0.16}^{+0.18}$\tabularnewline
\hline 
All $y_{Z}$ bins, & 60 & 2.19 & $0.79\pm0.03$ \tabularnewline
weighted average &  & 2.46 & $1.12\pm0.07$\tabularnewline
\hline 
\end{tabular}

\caption{The best-fit $\chi^{2}/N_{pt}$, central value and 68\% C.L. intervals
for $a_{Z}$ with fixed $C_{1,2,3}=\{b_{0},\ 1/2,\ b_{0}\}$ (upper
lines) and $\{2b_{0},\ 1/2,\ 2b_{0}\}$ (lower lines). \label{tab1}}
\end{table}

In method I, $a_Z$ is determined from the D\O\ data 
by minimization of a function
\ba
\chi^2(a_Z)=\sum_{i=1}^{N_{pt}}\left(\frac{D_i - 
\bar{T}_i(a_Z)}{s_i}\right)^2,
\label{chi2}
\ea
where $D_i$ are the data points; $\bar{T}_i(a_Z)$ 
are the theoretical predictions 
for fixed scale parameters $\{\bar{C}_1, \bar{C}_2, \bar{C}_3\}$;
$s_i$ are the uncorrelated experimental uncertainties; and
$N_{pt}$ is the number of points.

The dependence of $\chi^2$ on
$a_Z$ in three rapidity bins for two combinations of $\bar C_{1,2,3}$ 
is illustrated in Fig.~\ref{chivsa1}, and
the corresponding best-fit parameters are listed in Table~\ref{tab1}.
Electrons and muons are combined in the first two bins of rapidity,
$|y_Z|\leq 1$ and $1\leq|y_Z|\leq 2$. 
In both cases, the $\chi^2$ behavior is close to parabolic. The
locations of the $\chi^2$ minima are consistent in all three
bins. However, the quality of the fit is
unacceptable in the first two bins that have the smallest experimental
errors, with $\chi^2/N_{pt} \approx 3$. On the
other hand, the agreement is very good ($\chi^2/N_{pt} < 1$)
in the third bin, which has larger errors.

The weighted averages over all three bins are  $\bar a_{Z,\mbox{all y}}= 0.79\pm
0.03$ and $1.12\pm 0.07$ GeV$^{2}$ for the two scale combinations. The
location of the minimum is distinct from zero in both cases,
but its dependence on the scale parameters warrants further
investigation that we will now perform.

\subsubsection{Method II: computation with scale-parameter shifts}
\label{methodII}

To simplify the minimization when the scale parameters are
varied, we introduce a linearized approximation for the
covariance matrix of the type adopted for evaluating correlated
systematic effects in PDF fits~\cite{Pumplin:2001ct,Pumplin:2002vw}. For each scale
parameter $C_{\alpha}$, $\alpha=1,2,3$, we define a nuisance parameter
$\lambda_{\alpha}\equiv \log_2(C_\alpha/\bar C_\alpha)$ 
and compute the finite-difference derivatives of theory cross sections
\ba
\beta_{i \alpha} \equiv \frac{T_{i}(a_Z,\lambda_\alpha=+1) - T_{i}(a_Z,\lambda_\alpha=-1)} {2}, && \alpha=1,2,3; ~~~i=1,\dots,N_{pt} \,
\ea
over the interval $\lambda_\alpha = \pm 1$ corresponding to $\bar
C_\alpha/2 \leq C_\alpha \leq 2 \bar C_\alpha$. Variations of
$\lambda_\alpha$ introduce correlated shifts in theory cross sections
$T_i(a_Z,C_{1,2,3})$ with respect to the fixed-scale theory cross sections 
$T_i(a_Z,\bar C_{1,2,3})\equiv \bar T_i(a_Z)$. We can reasonably assume
that the probability distribution over each $\lambda_\alpha$ is similar to a
Gaussian one with a central value of 0 and half-width
$\sigma_\lambda$, taken to be the same for all $\lambda_\alpha$.  
The goodness-of-fit function is then defined as 
\ba
\chi^2(a_Z,\lambda_{1,2,3})=\sum_{i=1}^{N_{pt}}\left(\frac{D_i - \bar T_i(a_Z) -\sum_{\alpha=1}^3\beta_{\alpha i}
\lambda_\alpha}{s_i}\right)^2 +\sum_{\alpha=1}^3\frac{\lambda^2_\alpha}{\sigma_\lambda^2}.
\label{chi2shift}
\ea

The minimum with respect to $\lambda_{\alpha}$ can be found
algebraically for every $a_Z$ as~\cite{Pumplin:2001ct}
\begin{equation}
\min\chi^{2}=\chi^2(a_Z,\bar \lambda_\alpha)=\sum_{i,j}^{N_{pt}}(D_{i}-\bar T_{i}(a_Z))({\rm
  cov^{-1}})_{ij}(D_{j}-\bar T_{j}(a_Z)),\label{eq:chi2cov}
\end{equation}
containing the inverse of the covariance matrix, 
\begin{equation}
({\rm cov}^{-1})_{ij}=\left[\frac{\delta_{ij}}{s_{i}^{2}}-\sum_{\alpha,\beta=1}^3\frac{\beta_{i,\alpha}}{s_{i}^{2}}\mathcal{A}_{\alpha\beta}^{-1}\frac{\beta_{j,\beta}}{s_{j}^{2}}\right],
\end{equation}
and a matrix ${\mathcal A}$ given by
\begin{equation}
\mathcal{A_{\alpha\beta}}=\sigma_\lambda^2 \delta_{\alpha\beta}+\sum_{k=1}^{N_{pt}}\frac{{\beta}_{k,\alpha}\beta_{k,\beta}}{s_{k}^{2}}.\label{A}
\end{equation}
Eq.~(\ref{eq:chi2cov}) is essentially the standard 
$\chi^2$ function based on the covariance matrix in the
presence of the correlated shifts. For every $a_Z$, the nuisance 
parameters $\bar \lambda_\alpha$ that realize the $\chi^2$ minimum 
are also known, 
\begin{equation}
\bar{\lambda}_{\alpha}(a_Z)=\sum_{i=1}^{N_{pt}}\frac{D_{i}-\bar T_{i}(a_Z)}{s_{i}}\sum_{\delta=1}^3\mathcal{A}_{\alpha\delta}^{-1}\frac{\beta_{i,\delta}}{s_{i}}.\label{barlambda}
\end{equation}

\begin{figure}[tb]
\begin{center}
\includegraphics[height=6cm]{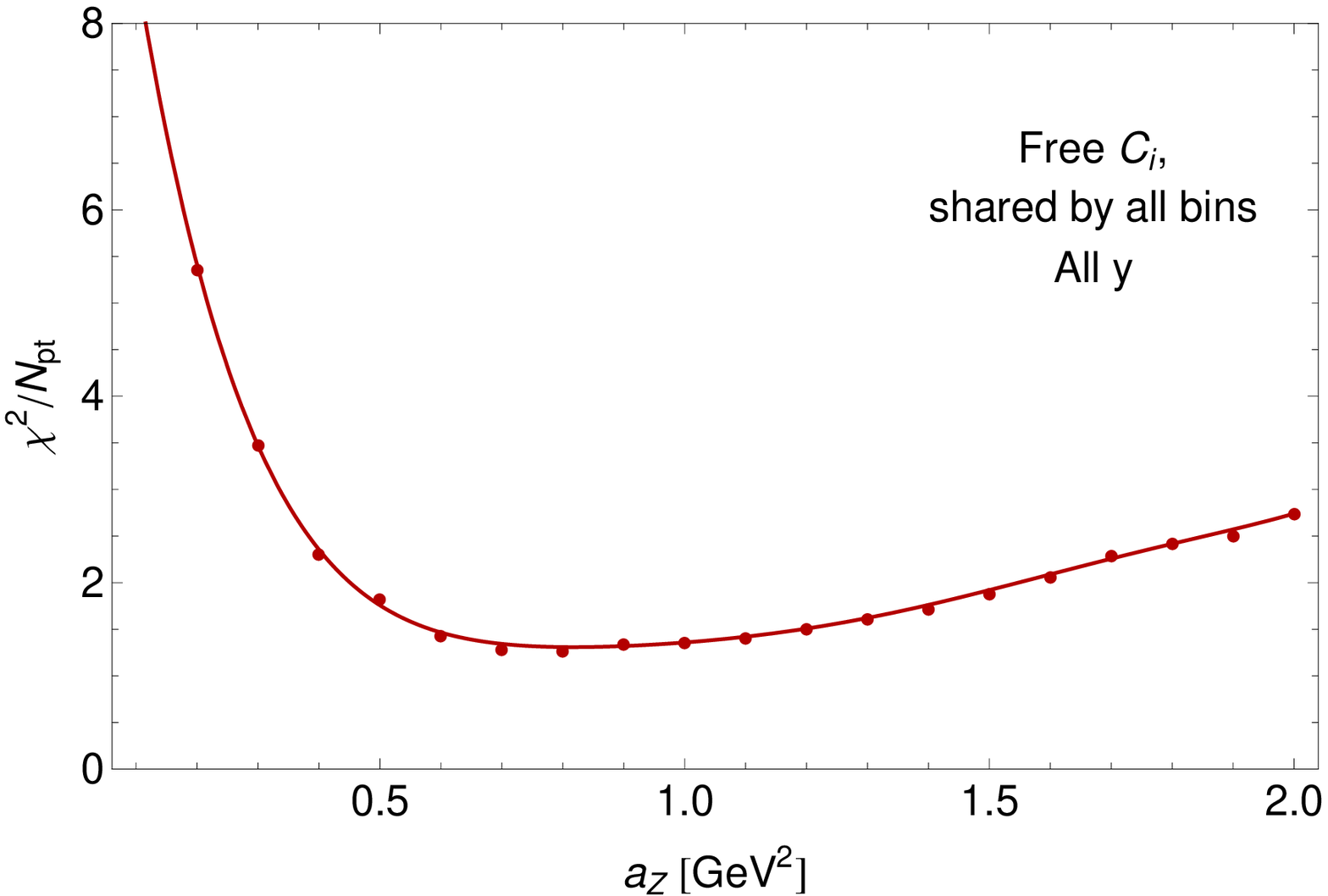}
\includegraphics[height=6cm]{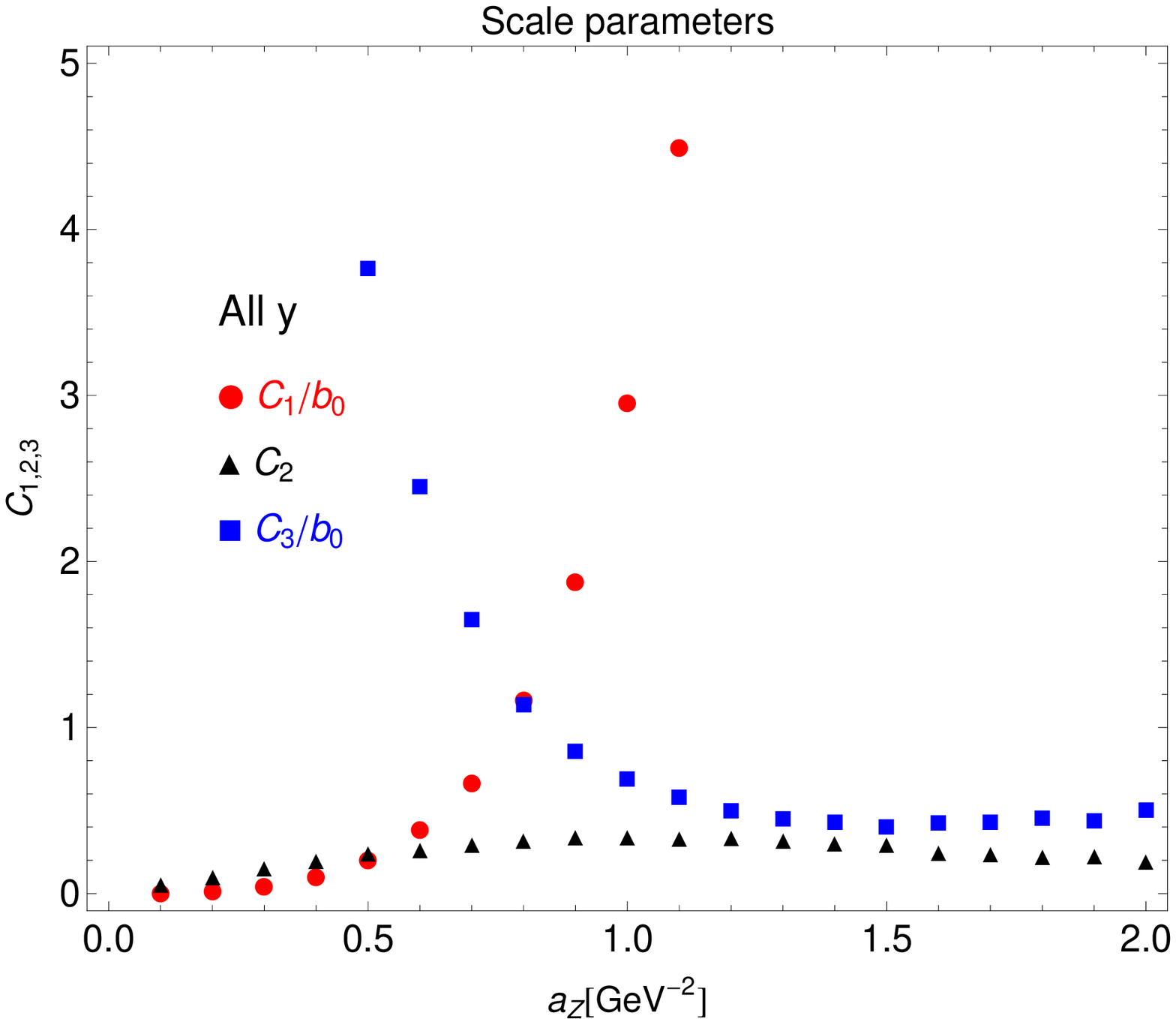}
\caption{$\chi^2/N_{pt}$ and scale parameters as a function of $a_Z$ for $\bar C_1=\bar
  C_3=2\ b_0, \bar C_2=1/2$. The scale parameters are shared
  across three $y_Z$ bins. 
\label{chivsa2}}
\end{center}
\end{figure}

\begin{table}
\begin{centering}
\begin{tabular}{|c|c|c|c|c|}
\hline 
\multicolumn{5}{|c|}{ Fit results for $\phi_{\eta}^{*}\leq0.1$}\tabularnewline
\hline 
\hline 
\multicolumn{5}{|c|}{$C_{1},$$C_{2},$ $C_{3}$ are shared by all $y_{Z}$ bins}\tabularnewline
\hline 
 & $N_{pt}$  & $\chi_{min}^{2}/N_{pt}$  & $\overline{a}_{Z}\pm\delta a_{Z}\ (\mbox{GeV}^{2})$  & Best-fit $C_{1,2,3}$\tabularnewline
\hline 
All $y_{Z}$ bins & 60 & 1.29 & $0.82_{-0.12}^{+0.34}$ & 1.4, 0.33, 1.23\tabularnewline
 &  & 1.31 & $0.82_{-0.11}^{+0.22}$ & 1.42, 0.33, 1.23\tabularnewline
\hline 
\hline 
\multicolumn{5}{|c|}{$C_{1},$$C_{2},$ $C_{3}$ are independent in each $y_{Z}$ bin}\tabularnewline
\hline 
 & $N_{pt}$  & $\chi_{min}^{2}/N_{pt}$  & $\overline{a}_{Z}\pm\delta a_{Z}\ (\mbox{GeV}^{2})$  & Best-fit $C_{1,2,3}$\tabularnewline
\hline 
$|y_{Z}|\leq1$, ~~$e+\mu$  & 24 & 1.0 & $0.56_{-0.02}^{+0.95}$ & 0.21, 0.18, 7.56\tabularnewline
 &  & 1.16 & $0.85_{-0.15}^{+0.3}$ & 1.47, 0.3, 1.46\tabularnewline
\hline 
$1\leq|y_{Z}|\leq2$,~~ $e+\mu$  & 24 & 1.48 & $1.22_{-0.36}^{+0.27}$ & 18, 0.58,0.1\tabularnewline
 &  & 1.70 & $0.79_{-0.1}^{+0.2}$ & 1.69, 0.37, 0.77\tabularnewline
\hline 
$|y_{Z}|\geq2$, ~~$e$ & 12 & - & - & -\tabularnewline
 &  & 0.59 & $0.99_{-0.31}^{+0.99}$ & 1.74, 0.48, 2.12\tabularnewline
\hline 
Weighted average & 60 &  & $0.97\pm0.25$ & \tabularnewline
of all bins &  &  & $0.82\pm0.12$ & \tabularnewline
\hline 
\end{tabular}
\par\end{centering}

\caption{The best-fit $\chi^{2}/N_{pt}$, central value and 68\% C.L. intervals
for $a_{Z}$, and best-fit $C_{1,2,3}$ for $1/\sigma_{\lambda}=0$
(upper rows in each section) and $1$ (lower rows). \label{tab2}}
\end{table}

Based on this representation for $\chi^2$ (designated as ``fitting
method II''), we explored the impact of the scale dependence on the
constraint on $a_Z$. Even if the scales are varied, 
data prefer a nonzero nonperturbative Gaussian smearing of about the
same magnitude as in method~I.

In the simplest possible case, 
the $C_{1,2,3}$ parameters are independent of the rapidity 
or other kinematic parameters and shared by all $e$ and $\mu$ bins. 
In this case, variations of the scale
parameters reduce $\chi^2/N_{pt}$ to about 1.3, {\it i.e.} the fit is
better than for the fixed scale combinations discussed above.
We focus on the case when the central scale parameters are  
$\bar C_1=\bar C_3=2\ b_0, C_2=1/2$, although the conclusions remain
the same for other choices.

The plots of $\chi^2/N_{pt}$ vs. $a_Z$ and optimal $C_1/b_0$, $C_2$,
$C_3/b_0$ vs. $a_Z$, derived from the optimal $\lambda_{\alpha}$
parameters in Eq.~(\ref{barlambda}), are shown in Fig.~\ref{chivsa2}.
The $\chi^2$ dependence on $a_Z$ becomes asymmetric when the scale
shifts are allowed, with the large-$a_Z$ branch being flattened out in
contrast to the small-$a_Z$ one that remains steeply growing. 
From the right inset,
we see that the optimal $C_1$ and $C_3$ are monotonously
increasing and decreasing as functions of $a_Z$, respectively. In the
vicinity of the minimum, $C_1$ and $C_3$ are of about the same
magnitude at $(1.2-1.5) b_0$. Very small or large $a_Z$ can be obtained 
only by taking $C_1$ and $C_3$ to be
uncomfortably far from unity. In contrast, the
optimal $C_2$ parameter is generally in the range 0.3-0.5 and has
weaker dependence on $a_Z$. 

The values of $\chi^2/N_{pt}$, $a_Z$, and $C_{1,2,3}$
parameters at the minimum are reported 
in the upper portion of Table~\ref{tab2}. 
When the $C_{1,2,3}$ parameters are shared by all bins,
the fit is relatively insensitive to the confidence level assigned to
the variations $\lambda_\alpha \pm 1$, controlled by 
the parameter $\sigma_\lambda$ in Eq.~(\ref{chi2shift}). In Table~\ref{tab1}, the
upper rows in each section correspond to the fit without a constraint 
on the $\lambda$ parameters, {\it i.e.}, for
$1/\sigma_\lambda=0$. The lower rows are for assigning a
68\% probability to the $-1\leq \lambda_\alpha\leq 1$ intervals, corresponding 
to $1/\sigma_\lambda=1$.

For the shared $C_{1,2,3}$, the outcomes
of the fits with $1/\sigma_\lambda=0$ and 1 are very similar, apart
from the uncertainty on the $a_Z$ parameter, which is increased when the
$\lambda_\alpha$ variations are totally free. [The asymmetric 68\% C.L.
uncertainties are computed according to Eq.~(\ref{aZmin_max})]. 

\begin{figure}[tb]
\begin{center}
\includegraphics[width=0.49\textwidth]{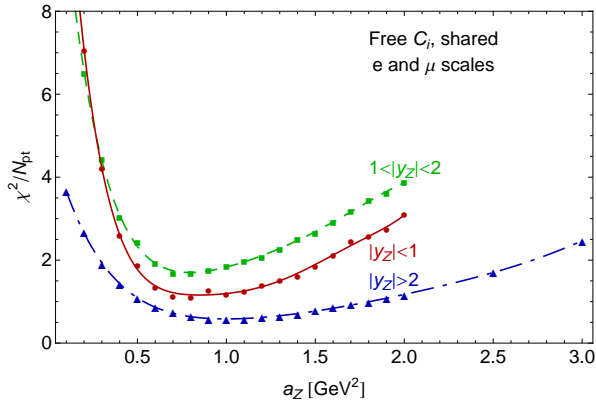}
\caption{$\chi^2/N_{pt}$ as a function of $a_Z$ for $\bar C_1=\bar
  C_3=2\ b_0, \bar C_2=1/2$. The scale parameters are independent in each
  $y_Z$ bin. 
\label{chivsa3}}
\end{center}
\end{figure}

In contrast, when the scale parameters are taken to be independent in
each $y_Z$ bin (but still shared between the electron and muon
samples), only the case of $\sigma_\lambda=1$ results in
an acceptable fit in all three $y_Z$ bins. The best-fit parameters 
for this case are listed in the lower part of
Table~\ref{tab2}. When the scale shifts were arbitrary
($1/\sigma_\lambda=0$, upper lines), the fits were underconstrained and produced
inconsistent $a_Z$ values and large scale shifts 
in all three bins, especially in the third bin that is not shown for
this reason.
On the other hand, for $\sigma_\lambda=1$
(lower lines), the three fits converged well and rendered compatible $a_Z$ values. 
The $\chi^2/N_{pt}$
vs. $a_Z$ dependence for this case is illustrated in
Fig.~\ref{chivsa3}, where the minima are neatly aligned
in the three bins. 
The fit to the second bin is generally worse than for the other two,
suggesting possible rapidity dependence of $a_Z$. The scale dependence
in each bin is qualitatively similar to that in the right inset of 
Fig.~\ref{chivsa2}.

\begin{figure}
\begin{center}
\hspace{-0.9cm}
\includegraphics[width=12cm, angle=0]{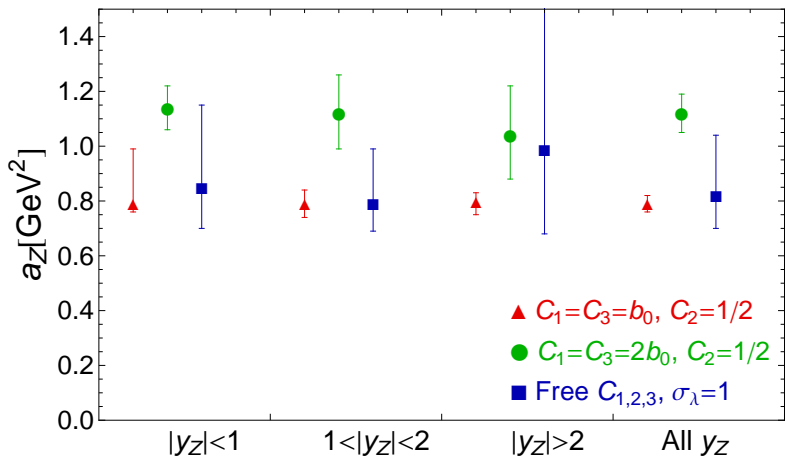}
\caption{68\% C.L. ranges for $a_Z$ in individual $y_Z$ bins and in
  all bins.\label{a1P}}
\end{center}
\end{figure}

Even when $C_{1,2,3}$ are independent in each $y_Z$ bin,
 by averaging the $a_Z$ values over three bins, we obtain the
$\bar a_Z=0.8-0.9\mbox{ GeV}^2$ in the last section of
 Table~\ref{tab2} that is essentially the same as in the case 
when $C_{1,2,3}$ are shared by all bins.
The findings in Tables~\ref{tab1} and \ref{tab2} 
are recapitulated in Fig.~\ref{a1P},
showing the 68\% C.L. intervals in the fits with fixed
$C_{1,2,3}={b_0,1/2,b_0}$ and ${2 b_0, 1/2, 2 b_0}$, 
as well as the fit with varied $C_{1,2,3}$ and $\sigma_\lambda=1$. 
All fits consistently yield $a_Z$ values that are at least 5$\sigma$ from zero. 

\section{Implications for the $W$ mass measurement 
and LHC \label{sec:Discussion}}
The previous sections demonstrated that the  $\phi^*_\eta$ 
distributions in $Z/\gamma^*$ production are sensitive to several QCD effects. 
Depending on the $\phi^*_\eta$ range, hard or soft QCD emissions can be studied. 
The nonperturbative power corrections in QCD can be determined at 
$\phi^*_\eta\leq 0.1$, provided that the dependence on resummation scales
is controlled.  

To distinguish between various contributing effects, new developments 
in the Collins-Soper-Sterman resummation formalism were necessitated. 
The computer code \textsc{ResBos} includes all such effects relevant
for computation of resummed differential distributions of lepton
pairs. New components of the theoretical framework implemented in
\textsc{ResBos} were reviewed in Sec.~\ref{sec:Overview}. In the
large-$\phi_\eta^*$ region dominated by hard emissions, the two-loop
fixed-order contributions implemented in \textsc{ResBos} show good
agreement with the D\O~data when the renormalization/factorization
scale $C_4 Q$ for hard emissions is set to be close to
$Q/2$.\footnote{In this region, a three-loop correction must be
  computed in the future to reach NNLO accuracy 
in $\alpha_s$.} 

In the resummed $W$ piece dominating at small $\phi^*_\eta$, 
we include 2-loop perturbative coefficients in the resummed $W$ term
by using the exact formulas for the ${\cal A}$ and ${\cal B}$
coefficients and a numerical estimate for the small ${\cal O}(\alpha_s^2)$
contribution $\delta C^{(2)}$ to the Wilson coefficient functions. 
We also fully include, up to ${\cal O}(\alpha_s^2)$,
the dependence on resummation scale parameters $C_1$ and $C_2$, (see Secs.~\ref{sec:WpertCanonical} and \ref{sec:WpertAS}).
Matching corrections and final-state electroweak contributions were
implemented and investigated in order to understand their
non-negligible impact on the cross sections. 
Finally, we implemented a form factor $\widetilde{W}_{NP}(b,Q)$ 
describing soft nonperturbative emissions at transverse positions $b\gtrsim 1\mbox{ GeV}^{-1}$ in 
the context of a two-parameter $b_*$ model~\cite{Konychev:2005iy},  cf. 
Sec.~\ref{sec:WNP}. 

With this setup, we performed a study of the small-$\phi^*_\eta$
region at the D\O\ Run-2 
with the goal to determine the range of plausible nonperturbative
contributions.  
We found that, to describe Drell-Yan dilepton production with the
invariant mass $70\leq Q \leq 110$ GeV, it suffices to use a
simplified nonperturbative form factor that retains only  
a leading power correction,
$\widetilde{W}_{NP}(b,Q=M_Z)=\exp\left(-b^2 a_Z\right)$.  
The power correction modifies the shape of $d\sigma/d\phi^*_\eta$ in a pattern 
distinct from variations due to the dependence on the resummation scales
$C_1/b$, $C_2 Q$, and $C_3/b$  
in the leading-power term $\widetilde{W}^{pert}$, 
see Figs.~\ref{scaleC2E}, \ref{scaleC2M}, and \ref{fig:aZkc1}. For
various fixed combinations  
of scale parameters $C_{1,2,3}$, or when the scale parameters were varied, 
the fits require nonzero $a_Z$ values that were summarized in
Tables~\ref{tab1} and \ref{tab2}. For example, when  
the variations in the scales $C_{1,2,3}$ were incorporated as shared
free parameters in all rapidity bins  
using a correlation matrix, we obtained
$a_Z=0.82^{+0.22}_{-0.11}\mbox{ GeV}^2$ at 68\% C.L.,  
cf. Table~\ref{tab2}, consistently with the other tried methods.
The estimate of the 68\% C.L. uncertainty including the scale
dependence indicates clear preference for a non-zero $a_Z$, without
appreciable rapidity dependence.

The magnitude of $a_Z$ depends on the
resummation scales, but allowing the scales to vary increases the
probability for having larger, not smaller $a_Z$. The best-fit $a_Z$ is also
correlated with $b_{max}$, which controls the upper boundary of the
$b$ range where the exact perturbative approximation
for $\widetilde{W}^{pert}(b,Q,y_Z)$ is used. Using $b_{max}=1.5\mbox{ GeV}^{-1}$
in this study, we obtain $a(b,Q) \approx 0.8\mbox{ GeV}^2$ at $Q=M_Z$, 
which is consistent with the value obtained with the other
$\widetilde{W}_{NP}$ forms maximally preserving the perturbative contribution~\cite{Qiu:2000hf,Tafat:2001in,Kulesza:2002rh,Konychev:2005iy}. 
The dependence on $b_{max}$ weakens at $b_{max}$
above $1\mbox{ GeV}^{-1}$, and even larger $a_Z$ values are preferred 
for $b_{max}$ below $1\mbox{ GeV}^{-1}$, cf. Fig.~2 in~\cite{Konychev:2005iy}. 
The fitted data was corrected for the effects of final-state NLO QED radiation.
In the fitted region $\phi^*_\eta < 0.1$, the uncertainty due to the
matching of the resummed and finite-order terms was shown to be negligible.

The nonperturbative form factor at other $\sqrt{s}$ and $Q$ values 
can be predicted using the relations in Sec.~\ref{sec:WNP}. 
This is possible because the dominant part 
of $\widetilde{W}_{NP}$ is associated with the soft factor 
$\exp\left(-S(b,Q)\right)$ 
which does not depend on $\sqrt{s}$ or the types of the incident
hadrons. It is argued in Sec.~\ref{sec:WNP} that the $\widetilde{W}_{NP}$
factors are identical within the 68\% C.L. error in central-rapidity $Z$ and $W$
production at the same $\sqrt{s}$. The same $a_Z$ value that we determined 
can be readily applied to predict $W$ boson differential
distributions at the Tevatron Run-2, 
or, with appropriate modifications proportional to
$\ln(Q)$ and $\ln(s)$, in other kinematical ranges, cf. Eq.~(\ref{WNPQ}).

The resummation calculation employed in this analysis can be
reproduced using the \textsc{ResBos-P}~code~\cite{RESBOSP} and input tables
~\cite{RESBOSgridsMG} available at the ``$Q_T$ resummation portal at Michigan
State University''. The central input tables are provided for 
$a_Z=1.12\pm 0.07 \mbox{ GeV}^2$, $C_1=C_3=2 b_0$, $C_2=1/2$, and
central CT10 NNLO PDF. In addition, the distribution includes 
\textsc{ResBos} tables corresponding to the best-fit resummed 
parameters and CT10 NNLO PDF eigenvector sets. Finally, for a detailed
exploration of the low-$\phi^*_\eta$ region, the distribution includes
tables for $a_Z$ in the interval $0.5 - 1.7\mbox{ GeV}^2$ 
with step $0.1\mbox{ GeV}^2$ using the central PDF,
and, to study scale dependence, 7 \textsc{ResBos} grids for 
the central $a_{Z,central}=1.12$ GeV$^{2}$, and the scale parameters 
$C_1={b_0,4b_0}$, $C_2={1/4,1}$, and $C_3={b_0,4b_0}$.

\begin{figure}
\begin{center}
\includegraphics[height=8cm, angle=0]{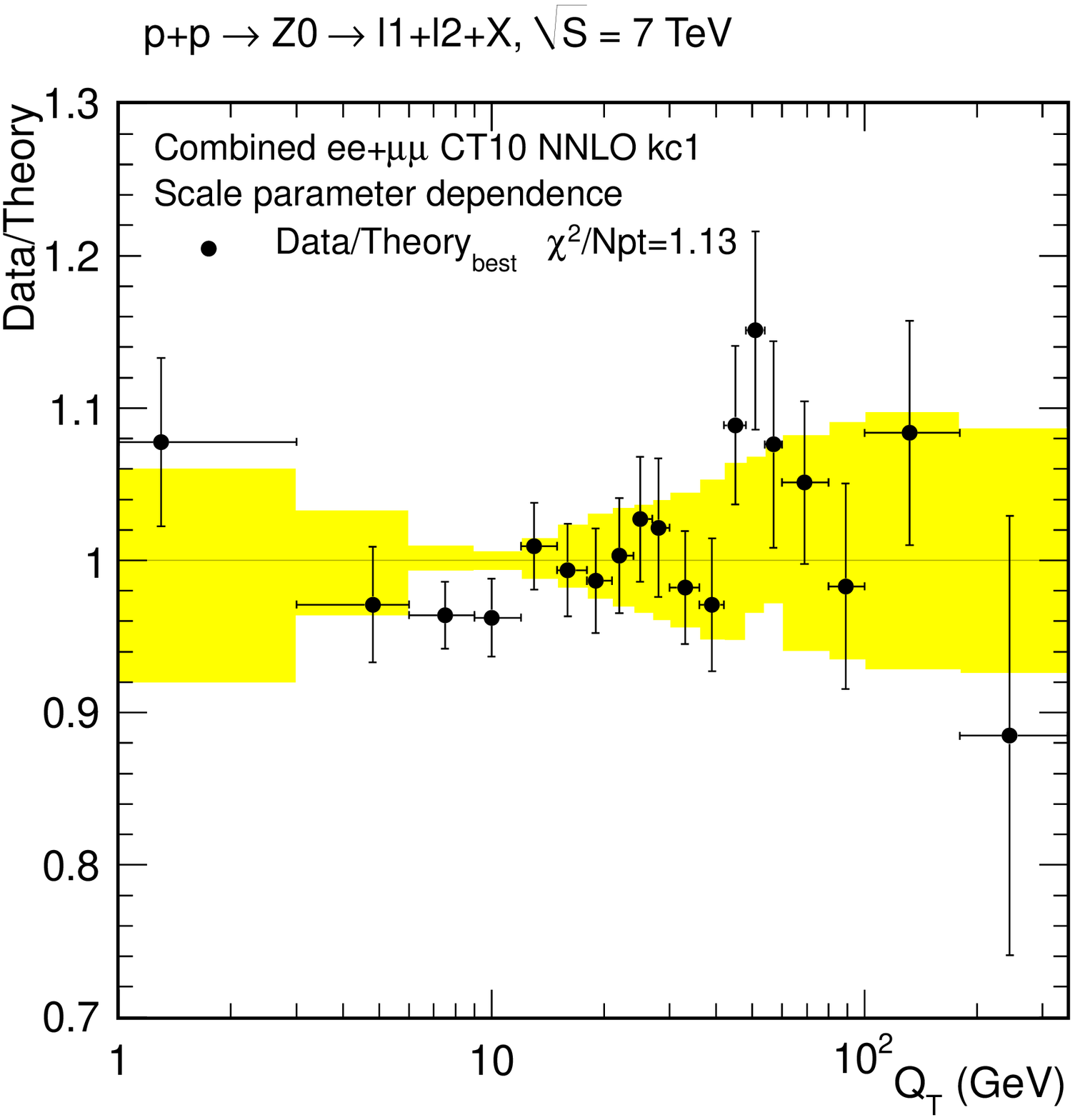}
\includegraphics[height=8cm, angle=0]{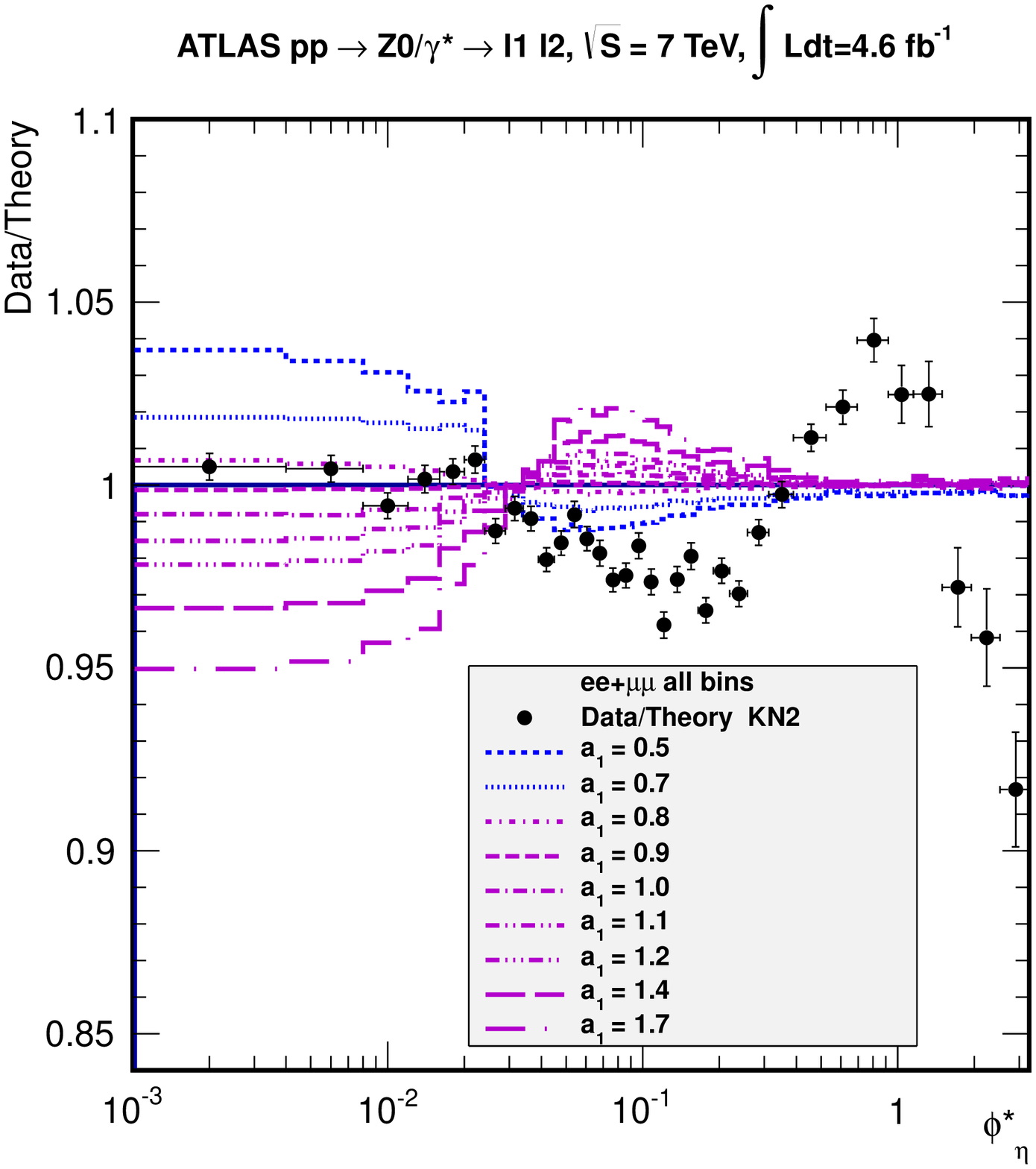}
\caption{Data vs. theory ratios for the $Q_T$ distribution by ATLAS 7
  TeV, $35-40 \mbox{ pb}^{-1}$~\cite{Aad:2011gj} and $\phi^*_\eta$ distribution  ATLAS 7
  TeV, $4.6 \mbox{ fb}^{-1}$~\cite{Aad:2012wfa} \label{fig:scaleLHC}}
\end{center}
\end{figure}

As an example of a phenomenological application,
Fig.~\ref{fig:scaleLHC} compares the \textsc{ResBos} predictions 
with the ATLAS data~\cite{Aad:2011gj,Aad:2012wfa} on Drell-Yan pair production near 
the $Z$ boson resonance peak at $\sqrt{s}=7$ TeV. The figure shows 
ratios of data to theory cross sections. 
The left subfigure
shows the $Q_T$ distribution for $35-40 \mbox{ pb}^{-1}$, compared to
the \textsc{ResBos} prediction with $a_Z=1.1$ GeV$^2$, $C_1=C_3=2\ b_0, C_2=1/2$. The yellow band indicates
variations in the cross section due to the scales in the range $C_1={b_0,4b_0}$, $C_2={1/4,1}$, and $C_3={b_0,4b_0}$. 
In the case of $Q_T$ distribution, we obtain
good agreement between theory and data and in the intermediate/small $Q_T$ region  
the theoretical uncertainty due to $C_{1,2,3}$ scale parameters 
is reduced compared to the study of Ref.~\cite{Banfi:2012du}.

The right subfigure shows the ratio of the more recent
$\phi^*_\eta$ distribution to the
central theory prediction based on our default parametrization 
at much higher level of accuracy. Here, a \textsc{ResBos} prediction 
based on the BLNY parametrization has shown better agreement with the data than other available codes and was 
used for event simulation during the ATLAS analysis. A comparable, although somewhat worse agreement 
is realized by the GNW parametrization, which was not used at any stage by ATLAS. 
The right subfigure shows several curves for the default $C_{1,2,3}$ choice 
and $a_Z$ in the range $0.5-1.7\mbox{ GeV}^2$. It is clear that the 
ATLAS $\phi^*_\eta$ data is sensitive to $a_Z$ as well as $b_{max}$ 
and can possibly discriminate subleading power contributions to the nonperturbative 
form factor $\widetilde{W}_{NP}(b,Q)$ proportional to $b^4$ and 
beyond. We provide sets of updated \textsc{ResBos} grids for the 
LHC kinematics that can be used for future improvements in the 
nonperturbative model.

\section{Conclusions}

In our analysis we have shown that a significant nonperturbative Gaussian smearing 
is necessary to describe features of the low $\phi^*_{\eta}$ spectrum. 
A non-zero NP function is present even if all the perturbative 
scale parameters of the CSS formalism are varied. 
Values of $a_Z$ smaller than $0.5$ GeV$^2$ are disfavoured 
by the fit to the recent D\O~data, as demonstrated in Sec.~\ref{sec:numerical-results}.
The dependence of the $d\sigma/d\phi_\eta^*$ on various factors was recently examined in~\cite{Banfi:2011dm}, and it was observed that 
the dependence of  on the nonperturbative contributions could not
be reliably separated from the dependence on the perturbative QCD scales. To go beyond the analysis 
of Ref.~\cite{Banfi:2011dm}, we carried out  a quantitative fit to the $\phi_\eta^*$ data of D\O~, 
in which we implemented the dependence on the soft resummation scales to NNLO, 
cf. Sec.~\ref{sec:WpertAS}. We found that the small-$\phi^*_{\eta}$ spectrum 
cannot be fully described by employing perturbative scale variations only.
From the characteristic suppression of the production rate 
at very small $\phi_\eta^*$, or very small $Q_T/Q$, 
we established the magnitude of the nonperturbative effects.

The resummed predictions based on the new nonperturbative form are implemented in the 
\textsc{ResBos} code. It will be of particular interest to explore the constraining power 
of the new forthcoming LHC data for $Z$ and $W$ production  
at a variety of $\sqrt{s}$, boson's invariant masses, and rapidities. 
Precise measurements of hadronic cross sections at small $Q_T$
will verify the TMD formalism for QCD factorization and shed light on
the nonperturbative QCD dynamics. These developments will depend on
consistent combination of NNLO QCD and NLO electroweak effects 
and reduction of perturbative scale dependence 
in QCD predictions for $Q_T$ distributions.

\section*{Acknowledgments}
M.G. and P.N. would like to thank Mika Vesterinen for useful
discussion and correspondence, Ted Rogers for his comments on the
manuscript, Rafael Lopes de S\`a, John Hobbs, and
C.-P. Yuan for stimulating discussions. M.G. thanks
George Sterman for the hospitality at Stony Brook University during
the 2012 ``D\O\ $W$ mass'' workshop. M.G. also would like to thank Kostas
Theofilatos for discussions on future  prospects of the $Q_T$
distribution measurement by the CMS collaboration at the LHC. This
work is supported by the U.S. DOE Early Career Research Award
DE-SC0003870 and by the Lightner Sams Foundation.

\bibliographystyle{h-elsevier3}


\end{document}